\documentstyle[12pt,eqsecnum,aps,epsf]{revtex}

\def\){\right)} 
\def\({\left(} 
\def\]{\right]} 
\def\[{\left[}
\def\tr{\mbox{tr}}
\def\str{\mbox{str}}

\newcommand{\beq}{\begin{eqnarray}}
\newcommand{\eeq}{\end{eqnarray}}
\newcommand{\mcal}[1]{{\mathcal #1}}

\newcommand{\makefig}[4]{\begin{figure}
                           \centerline{\epsfysize=#3 in \epsfbox{#2}} 
                           \caption{#4} \label{#1} 
                         \end{figure}}

\begin{document}
\setcounter{page}{0}
\title{ \vspace{3cm}
Physical Results from Unphysical Simulations \\
\vspace{1cm}
}
\author{Stephen Sharpe\thanks{sharpe@phys.washington.edu}
 and Noam Shoresh\thanks{shoresh@phys.washington.edu} \\
}
\address{Physics Department, Box 351560, University of Washington,
Seattle, WA  98195, USA}

\maketitle
\vspace{2cm}
\begin{abstract}
We calculate various properties of pseudoscalar mesons
in partially quenched QCD using chiral perturbation theory
through next-to-leading order. 
Our results can be used to extrapolate to QCD from
partially quenched simulations, as long as the latter
use three light dynamical quarks.
In other words, one can use
unphysical simulations to extract physical quantities---in this
case the quark masses, meson decay constants, 
and the Gasser-Leutwyler parameters $L_4-L_8$.
Our proposal for determining $L_7$ makes explicit use of
an unphysical (yet measurable) effect of partially quenched theories,
namely the double-pole that appears in certain two-point 
correlation functions.
Most of our calculations are done for sea quarks having
up to three different masses, except for our result for $L_7$,
which is derived for degenerate sea quarks.
\end{abstract}

\vfill
\thispagestyle{empty}
\noindent UW/PT 00-10

\noindent hep-lat/0006017 \hfill 

\tighten

\section{Introduction}
\label{sec:introduction}
A major obstacle to direct simulations of lattice QCD
is the difficulty in simulating with light dynamical quarks.
In particular,
the up and down quarks must be reached by a chiral extrapolation.
In present simulations one must do this extrapolation from roughly
$m_s/2$, where $m_s$ is the physical strange quark mass.
This is far from the light quark masses 
($\overline m= (m_u+m_d)/2 \approx m_s/25$).

The aim of this paper is to provide formulae which can aid in this
extrapolation. 
To do this we use chiral perturbation theory (ChPT) 
at next-to-leading order (NLO). 
The parameters of the chiral Lagrangian that enter at this order 
are the Gasser-Leutwyler (GL) coefficients, $L_1-L_{10}$.
An alternative way to view the extrapolation to QCD is that,
by fitting numerical results in a region where quark masses
are considerably larger than the physical light quarks,
but small enough that NLO chiral perturbation theory is 
sufficiently accurate, one determines the relevant $L_i$.
These are physical parameters of QCD, 
governing many different physical properties 
(e.g. pion masses {\em and} scattering amplitudes).
With the $L_i$ in hand, one can then extrapolate to QCD,
and, in particular, determine the physical light quark masses.
For example, as has been stressed in Refs.~\cite{CKN,ShSh},
determining the combination $2 L_8-L_5$ with only moderate accuracy
might allow one to rule out the interesting possibility that $m_u=0$.
The accuracy of extrapolation depends, of course, 
on the reliability of NLO chiral perturbation theory.
This can be studied by seeing how well the numerical
data fit the expected forms, including the curvature
predicted by chiral logarithms.

\makefig{fig:PQ}{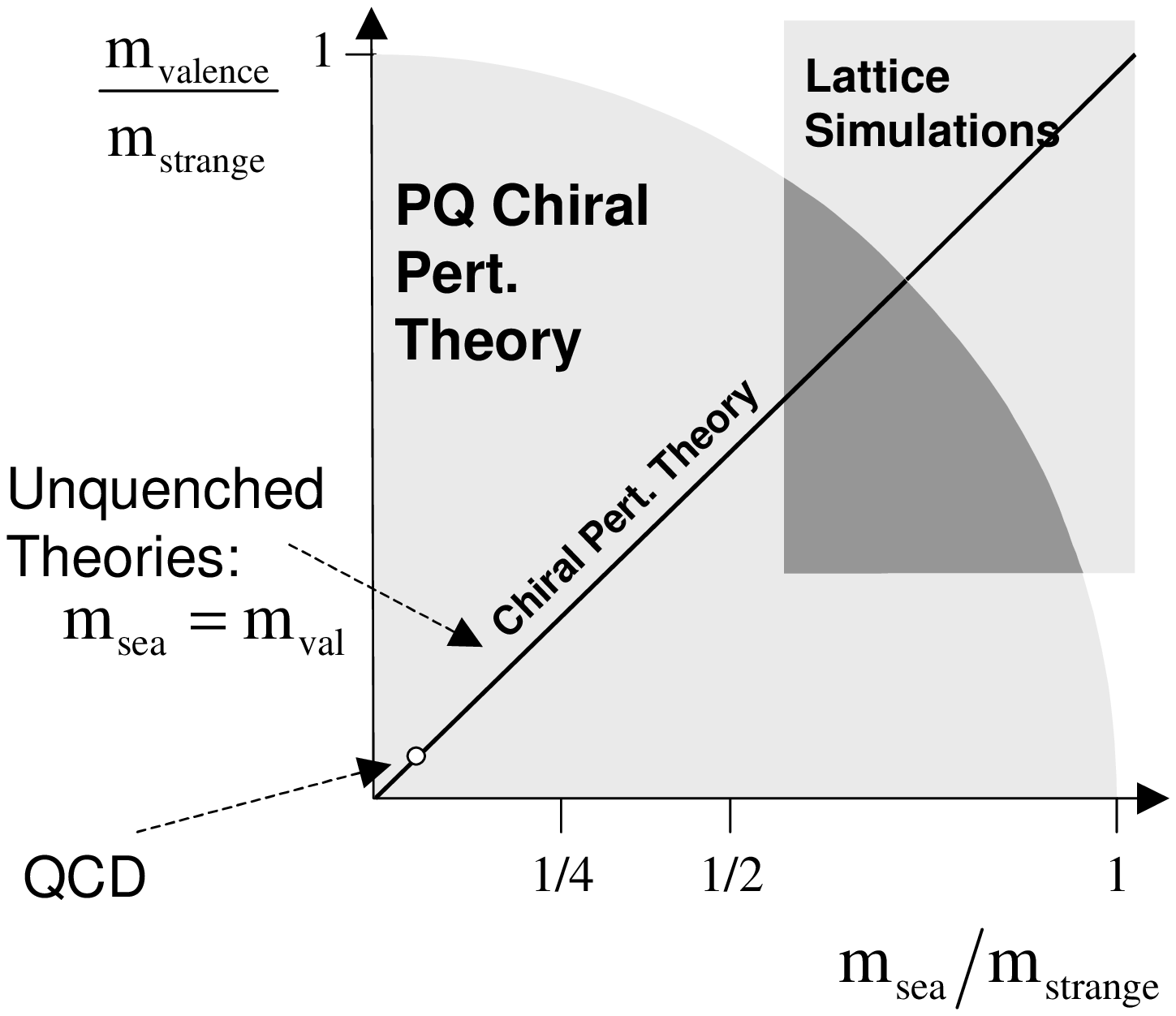}{2.5}{Schematic representation of the
space of PQ theories for ``light'' quarks (defined as lighter
than the physical strange quark mass). The approximate range of present
simulations is shown. The shape of this region is determined by
the fact that critical slowing down is less severe for valence
quarks than for sea quarks.}

An observation of practical importance is that
one can make use of partially quenched (PQ) simulations to aid
in the extrapolation to QCD~\cite{ShSh,CKN}.
In partially quenched simulations, 
one changes the mass of the valence quarks (typically reducing them), 
while holding the dynamical (or ``sea'') quark masses fixed.
The situation is illustrated schematically in Fig.~\ref{fig:PQ}.
This leads one into a space of unphysical theories, from which one
might expect to obtain only qualitative information about QCD.
It turns out, however, that, if all quark masses (valence and sea) 
are small enough, one can use PQ theories to obtain quantitative information
about unquenched theories.
Since it is computationally less demanding to reduce valence quark masses,
PQ simulations are often used to obtain approximate information on QCD.
Our point here is that they can be used to obtain {\em exact} information
about QCD. 

This observation follows from the structure of chiral perturbation theory
(ChPT) generalized to partially quenched theories---PQChPT~\cite{BGPQ}.
The key point is that there is a subspace of quark masses
(corresponding to the diagonal line in Fig.~\ref{fig:PQ}) where PQChPT
is completely equivalent to chiral perturbation theory for unquenched,
QCD-like theories.
Since the quark mass dependence in PQChPT is explicit, as in ChPT, 
it follows that the parameters of the PQ chiral Lagrangian
(with 3 light sea quarks) are the same as those of QCD.
These parameters do, however, depend
on the number of sea quarks, $N$. 
This means that PQ simulations with $N=3$ light sea quarks, whatever their
precise masses, give information about the parameters of the
chiral Lagrangian of QCD.
On the other hand PQ simulations with $N=2$ or $N=4$, the values
used in most simulations to date, {\em do not} give direct information
about QCD, even after extrapolation. 

These comments motivate the calculation of the NLO chiral corrections
to physically interesting quantities in PQ theories. 
Some results already exist in the literature: those for
charged pion masses and decay constants with degenerate sea 
quarks~\cite{SSPQ,GLPQ}, 
heavy-light meson properties~\cite{SZ},
vector and tensor meson properties~\cite{ChowRey},
baryons masses at large $N_c$~\cite{ChowReybaryons},
and electroweak amplitudes~\cite{GLPQ}.
We provide here two new results.
First, charged pion masses and decay constants are considered for
non-degenerate sea quarks (having up to three different masses).
This completes the calculations of simple pion properties for
any theory that is likely to be simulated.
It allows one to determine $L_{4-6}$ and $L_8$.
Note that, although non-degenerate sea quarks are not necessary
in order to extract these GL coefficients, 
as has been stressed in Refs.~\cite{CKN,ShSh},
there is no drawback to using them.
Indeed, someone might prefer to extrapolate using more
``QCD-like'' simulations with 
two degenerate ``light'' dynamical quarks and one dynamical quark with
mass fixed close to the physical strange quark mass.
Our formulae apply for such a theory.

Our second new result concerns the GL coefficient $L_7$.
In QCD, this contributes only to neutral meson masses, and does so
proportional to $(m_s-\overline m)^2$.
To determine $L_7$ using meson masses from
unquenched simulations thus requires non-degenerate quarks.
At first sight, PQ simulations do not improve the situation:
neutral meson masses are still independent of $L_7$ when
the sea quarks are degenerate. We find, however, that one
can determine $L_7$ from the coefficient of the double-pole 
in the propagators of neutral valence mesons,
even for degenerate sea quarks.
This is a nice example of the utility of PQ theories.
Although the double-pole is itself an
indicator that these theories are unphysical, its effects can
nevertheless be measured in lattice simulations, and its inferred
coefficient turns out to be related to a physical quantity.

Throughout our calculations, we treat the $\eta'$ as a heavy particle,
and integrate it out, following Ref.~\cite{SSPQ}. 
This greatly simplifies the resulting expressions,
since it removes the dependence on additional $\eta'$ coupling constants.
It raises, however, two important issues.
First, is the $\eta'$ heavy enough that removing it is appropriate?
The answer depends on the number of light sea quarks, $N$, 
and the number of colors, $N_c$. 
For the physical values of these parameters, $N=N_c=3$,
we know from experiment that $M_{\eta'}\approx 1\;$GeV.
Since this is the scale at which chiral perturbation theory breaks down,
it is appropriate to integrate out the $\eta'$ for these theories.
We stress, however, that our formulae are only valid when the
dynamical quarks are light enough that all pseudo-Goldstone
bosons, including those composed only of sea quarks, 
satisfy $M_{PGB}^2\ll M_{\eta'}^2$.
For further discussion of this point, and
of the limitations of the approach taken here, see Ref.~\cite{GLPQ}.

The second issue is more technical. 
How does one integrate out the $\eta'$ in PQ theories?
In this paper we follow Ref.~\cite{SSPQ} and do this by hand, 
working only at tree level.
This is unsatisfactory, since, as we know from QCD, $\eta'$ loops
to all orders give contributions of the same order in chiral
perturbation theory. In other words, the $\eta'$ must be
integrated out non-perturbatively. We return to this issue
in a companion paper, where we demonstrate that the approach
adopted here is equivalent to integrating out the
$\eta'$ non-perturbatively in the PQ theory~\cite{ShShNP}.

This paper is organized as follows. In the following section we 
recall the formalism of PQ chiral perturbation theory, 
and explain our calculational procedure.
After presenting a simple form for the neutral meson propagator
in Sec.~\ref{sec:calc},
we give our results for charged pion properties
in Sec.~\ref{sec:res}. We analyze these results in Sec.~\ref{sec:anal},
paying particular attention to the convergence of the
chiral expansion and the importance of non-analytic terms.
In Sec.~\ref{sec:L7} we explain how to extract $L_7$ using PQ theories.
We end with some conclusions.
Two appendices deal with technical issues in the calculation of
the neutral meson propagator.

Some parts of this work have been reported previously in Ref.~\cite{ShSh}.

\section{Theoretical framework}
\label{sec:theory}

We consider partially quenched theories with the following quark
complement: $N_1+N_2+N_3=N$ sea quarks,
$N_i$ each of mass $m_i$; 
two valence quarks with masses $m_A$ and $m_B$;
and two corresponding ghost quarks with masses $m_{\tilde A}=m_A$ 
and $m_{\tilde B}=m_B$.
The ghosts are needed to cancel the determinant arising
from the valence quark functional integral~\cite{Morel}.
In the chiral limit, this theory has an $SU(N+2|2)_L\otimes
SU(N+2|2)_R$ symmetry group~\cite{BGPQ}.
If $N_1=N_2=N_3=1$, and $m_1=m_u$, $m_2=m_d$ and $m_3=m_s$,
then the sea-quark sector is QCD.
Generalizing to arbitrary numbers $N_i$ covers
most other theories that are likely to be simulated
in an effort to shed light on QCD.

An important property of PQ theories, which follows trivially from
their definition, is that the
sea-quark sector decouples from the valence sector.
To be precise, 
all correlation functions composed of only sea-quark fields
are identical to those in the unquenched sea-quark theory.
There is no ``back-reaction'' from the valence sector.
The same result must also hold for the low energy chiral Lagrangian 
describing the PQ theory: correlators of pseudo-Goldstone mesons
composed of sea quarks should be the same as in the chiral Lagrangian
describing the unquenched theory. 
This was shown to be true in Ref.~\cite{BGPQ}.
In practice, however, one might view all correlation functions that
are calculated as being those of valence quarks, 
and so it is more useful
to reformulate this property as follows.
When each of the valence quarks
is assigned a mass that is equal to one of the sea quark masses, 
sea and valence quarks become indistinguishable,
and the cancelations of
the ``doubled'' quark species against their ghost counterparts
trivially render the theory the same as an unquenched theory
containing only the ``original'' sea quarks.
That this is so was also shown in Ref.~\cite{BGPQ},
and it has important consequences in the following.

At low energies, the partially quenched chiral effective theory is 
expressed in terms of the fields~\cite{BGPQ}%
\footnote{The normalization of $\Phi_0$ is different from that used in
Ref.~\protect\cite{SSPQ}, although it agrees when $N=3$ as in QCD.}
\beq
\Sigma &=&\exp({2i\Phi}/{f})\,,\\
\Phi &=&\(\Phi_{ab}\)=\frac{1}{\sqrt{2}}\(\pi_{ab}\),
\ a,b\in\{A,B,1,2,\ldots,N,\tilde{A},\tilde{B}\}\,, \\
\Phi_0 &=&\str\, \Phi/\sqrt{N} \,,
\eeq
and the quantity
\beq
\chi =2 \mu m=2\mu \mbox{ diag}(m_A,m_B,\underbrace{m_1,\ldots,m_1}_{N_1},
\underbrace{m_2,\ldots,m_2}_{N_2},\underbrace{m_3,\ldots,m_3}_{N_3},m_A,m_B)\,.
\eeq
where $m$ is the quark mass matrix. In the following we use the
notation $\chi_A=2\mu m_A$, $\chi_1 = 2\mu m_1$, etc.
The constants $f$ and  $\mu$ are unknown parameters.
The fields $\pi_{ab}$ describe the pseudo-Goldstone particles
of the theory---we refer to them generically as mesons even though
some are fermionic.
Because of the anomaly,
arbitrary functions of the field $\Phi_0$ (the super-$\eta'$) 
can appear in the Lagrangian.

The partially quenched chiral effective Lagrangian is expanded in 
powers of $\epsilon^2\sim M^2/\Lambda^2\sim p^2/\Lambda^2$, 
where $M$ is a typical pseudoscalar meson mass, 
$p$ the momentum, and
$\Lambda\sim 1\;$GeV is the scale beyond which the theory breaks down.

The parts of  the Euclidean Lagrangian contributing
to meson masses and decay constants at one-loop order are:
\beq
\mcal{L}_{\rm LO}&=&\frac{f^2}{4}
\str\( \partial_\mu\Sigma\partial_\mu\Sigma^\dagger\)
-\frac{f^2}{4}\str\(\chi\Sigma^\dag+\Sigma\chi\)
+\alpha\partial_\mu\Phi_0\partial_\mu\Phi_0+m_0^2\Phi_0^2
\label{eq:chL1}\\
\mcal{L}_{\rm NLO,1}&=&L_4\, 
\str\(\partial_\mu\Sigma\partial_\mu\Sigma^\dagger\)
\str\(\chi\Sigma^\dag+\Sigma\chi\)
+L_5\,\str\left[\partial_\mu\Sigma\partial_\mu\Sigma^\dagger
\(\chi\Sigma^\dag+\Sigma\chi\)\right]\label{eq:chL2} \nonumber\\
& &\mbox{    }-L_6\left[\str\(\chi\Sigma^\dag+\Sigma\chi\)\right]^2
-L_8\,\str\(\chi\Sigma^\dag\chi\Sigma^\dag+\Sigma\chi\Sigma\chi\) \\
\mcal{L}_{\rm NLO,2}&=&-L'_7\left[\str\(\chi\Sigma^\dag-\Sigma\chi\)\right]^2
+v_1 \Phi_0^2 \str\( \partial_\mu\Sigma\partial_\mu\Sigma^\dagger\) 
\nonumber\\
& &\mbox{    }
+v_2 \Phi_0^2 \str\(\chi\Sigma^\dag+\Sigma\chi\)
\label{eq:chL3}
\eeq
The coefficients $\alpha$, $m_0$, $L_i$ and $v_i$ are 
further unknown parameters of the low energy theory.\footnote{%
We revert to the notation $\alpha$ of Ref.~\cite{BG}, rather than the
$\alpha_\Phi$ used in Ref.~\cite{SSPQ}.}
The $L_i$ depend, in general, on the renormalization scale.
The NLO Lagrangian is broken into two parts because 
flavor off-diagonal mesons
receive contributions only from $\mcal{L}_{\rm NL0,1}$.

At this point we can make clear the relationship between the
partially quenched chiral Lagrangian and that
describing low energy QCD.
The latter is obtained by setting $N=3$, and ``unquenching'' --- 
i.e. assigning $m_A$ and $m_B$ values from $\{m_1,m_2,m_3\}$.
It follows that the unknown coefficients in $\mcal{L}$ are, for $N=3$,
identical to those in the QCD chiral Lagrangian.
This shows that these constants also govern the chiral behavior
of PQ extensions of QCD.

In QCD, one can take a further step and ``integrate out'' the $\eta'$. 
This is appropriate since it is not a pseudo-Goldstone boson,
having $M_{\eta'}^2 \approx m_0^2 + O(m)\approx 1\;{\rm GeV}^2$. 
Technically, the matching between theories with and without the
$\eta'$ is non-perturbative, since loops involving the $\eta'$
are not suppressed by powers of $M^2$ or $p^2$.
Thus in the standard approach one simply writes down the
Lagrangian without the $\eta'$, and it has the same form as
Eqs.~(\ref{eq:chL1})-(\ref{eq:chL3}), 
except that $\Phi$ is traceless.
It follows that $\alpha$, $m_0$ and the $v_i$ are irrelevant,
and the only NLO coefficients are the $L_i$.
It is in fact in this theory that the $L_i$---the Gasser-Leutwyler
coefficients---are conventionally defined.

In previous work on PQQCD, the step of integrating out
the $\eta'$ has been done by hand, i.e. at the level of individual
diagrams rather than the Lagrangian~\cite{SSPQ}.
We summarize the procedure here---details will become
apparent in the following section.
\begin{itemize}
\item
Loop diagrams involving the $\eta'$ are dropped, since these
lead to shifts in the parameters $L_i$ which are automatically
included if we use the $L_i$ from the QCD Lagrangian without the $\eta'$.
\item
Couplings special to the $\eta'$, such as the $v_i$ in Eq.~(\ref{eq:chL3}),
are treated as small, of $O(\epsilon^2)$, and thus appear only at tree
level. The justification for this treatment is that these couplings
are suppressed by powers of $1/N_c$, in this case $1/N_c^2$.
\item
On the other hand, the parameter $m_0^2$
is treated non-perturbatively since it is known to be $\sim\Lambda^2$, 
despite the fact that it is proportional to $1/N_c$.
In particular, we treat $M^2/m_0^2$ as $O(\epsilon^2)$
(with $M$, as above, a typical meson mass).
For convenience, we also treat $\alpha$ non-perturbatively.
\end{itemize}
While this procedure may be accurate enough for phenomenological purposes,
it is theoretically unsatisfactory because the $\eta'$ 
should be integrated out non-perturbatively. 
As noted in the introduction, we will
address this concern in a separate paper~\cite{ShShNP}.
In particular, we will argue that the procedure adopted here
in fact leads to results that are equivalent to those obtained from
integrating out the $\eta'$ non-perturbatively.

We close this section by deriving a result needed in Sec.~\ref{sec:L7}.
We claimed above that discarding $\eta'$ loops allows us to write our
results in terms of the standard GL coefficients, i.e. those in an
effective Lagrangian without the $\eta'$. 
This is not quite correct.
Tree diagrams involving an intermediate $\eta'$, which we keep, lead to a
shift in $L_7$ proportional to $M^2/m_0^2$.
Thus we are not using the conventional $L_7$, but
rather that defined in the effective theory containing the $\eta'$,
which we denote $L'_7$ [as anticipated in Eq.~(\ref{eq:chL3})].
In order to express our results in terms of conventional parameters,
we need to relate $L'_7$ to $L_7$ within the approximations of our
procedure.

To determine this relation we need consider only the sea-quark sector,
and thus work with the conventional (unquenched) chiral Lagrangian.
The $\eta'$ field in this theory is the
restriction to the sea sector of the ``super-$\eta'$'' field $\Phi_0$. 
To make the $\eta'$ dependence explicit, 
we decompose $\Sigma$ into pseudo-Goldstone and $\eta'$ parts:
\beq
\Sigma= U \exp(\frac{ 2 i \Phi_0}{f \sqrt{N}}) \,;\qquad U\in SU(N) \,,
\eeq
and substitute into the chiral Lagrangian.
The result is the original form with $\Sigma \to U$ and $\Phi_0\to0$,
i.e. the standard QCD chiral Lagrangian,
plus the following terms
\beq
{\cal L}_{\Phi_0} &=&
 c_1 \Phi_0 + c_2 \Phi_0^2 + c_3 (\partial \Phi_0)^2 + O(\Phi_0^3)\\
c_1 &=& i f  \mbox{tr}(\chi U^\dagger - \chi U)/(2\sqrt{N}) + O(\chi^2) 
\\
c_2 &=& m_0^2 + O(\chi)  \\
c_3 &=& 1 + \alpha + O(\chi)
\eeq
Only the leading order terms in the 
chiral expansion of the coefficients are shown, 
since higher order terms give contributions to 
the conventional chiral Lagrangian of orders $\epsilon^6$ and higher, 
too high to effect the matching of the GL 
coefficients which appear at order $\epsilon^4$.
For the same reason, the $(\partial \Phi_0)^2$ term can be dropped.
Keeping the $\eta'$ in tree graphs amounts to doing the functional
integral over $\Phi_0$ keeping only linear and quadratic terms.
The result is a contribution to the
conventional chiral Lagrangian of order $\epsilon^4$ 
with the same form as the $L_7$ term:
\beq
-{f^2 \over 16 N m_0^2} [\mbox{tr}(\chi U - \chi U^\dagger)]^2 (1 + O(\chi))
\,.
\eeq
Thus we find, within our approximations, the relation
\beq
L_7 = L'_7-{f^2 \over 16 N m_0^2} \,.
\label{eq:L7shift}
\eeq

\section{Calculation}
\label{sec:calc}

Expanding $\mcal{L}_{LO}$ to quadratic order, we obtain the
meson propagators. For ``charged'' mesons (i.e.
flavor off-diagonal states) these are  
\beq
G^C_{ab}(p) \equiv 
\int d^4x e^{-i p\cdot x} \langle \pi_{ab}(x) \pi_{ba}(0)\rangle =
{\epsilon_b \over p^2 + (\chi_a + \chi_b)/2 } \quad (a\ne b) \,,
\label{eq:GC}
\eeq
where the signature vector is
\beq
\epsilon_a=\left\{
  \begin{array}{cl}
    1 & a\in\{A,B,1,2,3,\dots,N\}\\
    -1 & a\in\{\tilde{A},\tilde{B}\}
  \end{array}
\right. .
\eeq

The propagators for ``neutral'' (flavor-diagonal) mesons 
include the contributions of the super-$\eta'$ interactions.
A general expression has been given in Ref.~\cite{BGPQ},
but we find it convenient to use an alternative form.
The propagator is a matrix acting on the space of neutral meson fields,
$\pi_{aa}$, $a=1, N+4$. 
In app.~\ref{app:GNLO} we show that
\beq
G^N_{ab}&\equiv& 
\int d^4x e^{-i p\cdot x} \langle \pi_{aa}(x) \pi_{bb}(0)\rangle \nonumber\\
&=&\frac{\epsilon_a\delta_{ab}}{p^2+\chi_a}
-\frac{(m_0^2+\alpha p^2)/N}{\(p^2+\chi_a\)\(p^2+\chi_b\)}
\frac{\(p^2+\chi_1\)\(p^2+\chi_2\)\(p^2+\chi_3\)}
{(1+\alpha)\(p^2+M_{\pi_0}^2\)\(p^2+M_{\eta}^2\)\(p^2+M_{\eta'}^2\)}\;.
\label{eq:GNtree}
\eeq
Here the ${\pi_0}$, $\eta$ and ${\eta'}$ are  
neutral mesons in the sea-quark sector. For $N=3$ they are
the usual neutral mesons of QCD; for $N>3$ they are the
appropriate generalizations, as explained in the appendix.
Their masses are functions of the sea-quark masses, and of the $N_i$,
as given explicitly in Eqs.~(\ref{eq:chipi1})-(\ref{eq:chipi3}).

The neutral propagator shows explicitly the unphysical nature of
the PQ theory. For example, if $a=b=A$, the second term has a double-pole
at $p^2=-\chi_A$. These double-poles are absent, however, in the physical,
sea-quark, sector. This was shown in Ref.~\cite{BGPQ},
but is particularly transparent with our result.
For example if $a=b=1$  (or equivalently if $a=b=A$ and $\chi_A=\chi_1$),
then the $(p^2+\chi_1)$ in the numerator reduces the double-pole to
a single-pole. 

For calculations, it is preferable to rewrite the propagator as
a sum of (single or double) poles. 
For simplicity, we discuss the case when $a\ne b$ and $\chi_a\ne\chi_b$,
for then $G^N_{ab}$ has only single poles.
For uniformity of notation, we introduce the definitions
\beq
\chi_\pi = M_{\pi_0}^2 \,,\quad
\chi_\eta = M_{\eta}^2 \,,\quad
\chi_{\eta'} = M_{\eta'}^2 \,.\quad
\eeq
in terms of which
\beq
G_{ab}^N
&=&
- \frac{1}{N} \sum_{x=a,b,\pi_0,\eta,\eta'}
{ R_x \over p^2 + \chi_x } \qquad (a\ne b)\,,\\
R_x &=& {(m_0^2 - \alpha \chi_x)\prod_{i=1,3} (\chi_i-\chi_x) \over
(1+\alpha) \prod_{y\ne x} (\chi_y-\chi_x) } \,,
\label{eq:residue}
\eeq
where $x$ and $y$ run over $a,b,\pi,\eta,\eta'$.
The degenerate limit $\chi_a=\chi_b$ or $a=b$, 
which, in general, has double poles
is straightforward to obtain.

At this point we are in a position to integrate out the $\eta'$ by hand,
bearing in mind that this propagator appears in loops.
First, as discussed above, we drop the $\eta'$ pole.
Second, we expand the residues of the remaining
poles in powers of $\chi_x/m_0^2$, $x\ne\eta'$,
and drop all but the leading term.
This is justified since we use the neutral propagator in one-loop diagrams,
which already give NLO contributions.
The only exception is in our discussion of $L_7$ in Sec.~\ref{sec:L7},
where the neutral propagator appears at tree-level.
With these changes, the residues become
\beq
R_x = {\prod_{i=1,3} (\chi_i-\chi_x) \over
\prod_{y=a,b,\pi,\eta \atop y \ne x} (\chi_y-\chi_x) } \,.
\label{eq:finalresidue}
\eeq
Note that both $m_0^2$ and $\alpha$ disappear from the neutral propagator.

\section{NLO results}
\label{sec:res}

In this section we calculate the properties of a meson composed of
two valence quarks to NLO. We use dimensional regularization,
and subtract the poles following the conventions of Ref.~\cite{DGH}.

Its mass, $M_{AB}$, is obtained from the diagrams of Fig.~\ref{Mass}.
We find
\beq
M_{AB}^2&=&\frac{\chi_A+\chi_B}{2} \left(1
        + \delta^M_{\rm tree} + \delta^M_{\rm loop} \right) 
\label{eq:MABNLO}\\
\delta^M_{\rm tree}
&=&\frac{8N}{f^2}\(2 L_6-L_4\) \bar\chi
+\frac{4}{f^2}\(2L_8-L_5\) (\chi_A+\chi_B)
\label{eq:mpiNLO}\\
\delta^M_{\rm loop} 
&=& \frac{1}{16f^2\pi^2 N} \left\{R_A \chi_A \log\chi_A
+ R_B \chi_B \log\chi_B
+ R_\pi \chi_\pi \log\chi_\pi
+ R_\eta \chi_\eta \log\chi_\eta \right\}
\eeq
Here $\bar\chi$ is the average sea-quark mass,
\beq
\bar\chi = \frac{1}{N} \sum_{i=1,3} N_i \chi_i \,,
\eeq
and the residues are defined in Eq.~(\ref{eq:finalresidue}).
For concreteness we quote two explicit examples 
\beq
R_A &=& \frac{\(\chi_A-\chi_1\)\(\chi_A-\chi_2\)\(\chi_A-\chi_3\)}
{\(\chi_A-\chi_B\)\(\chi_A-\chi_\pi\)\(\chi_A-\chi_{\eta}\)}\,,
\nonumber\\
R_\pi &=& \frac{\(\chi_\pi-\chi_1\)\(\chi_\pi-\chi_2\)\(\chi_\pi-\chi_3\)}
{\(\chi_\pi-\chi_A\)\(\chi_\pi-\chi_B\)\(\chi_\pi-\chi_{\eta}\)} \,.
\nonumber
\eeq
$R_B$ is obtained from $R_A$ by interchanging $A$ and $B$,
and $R_\eta$ is obtained from $R_\pi$ by interchanging $\pi$ and $\eta$.

\makefig{Mass}{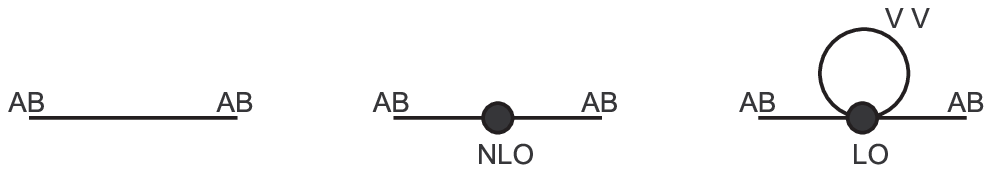}{1}{Diagrams contributing to $M_{AB}$. The
  letters next to the lines denote the flavor indices of the propagating
  mesons. ``VV'' stands for Valence-Valence meson with $V=A,B$. ``LO''
  and ``NLO'' describe the order of the vertex that makes the diagram
  contribute at 1 loop.}

The renormalization scale is implicit in the logarithms and
the $L_i$. Using the result 
\beq
\sum_{x=A,B,\pi,\eta} \chi_x R_x = (\chi_A+\chi_B-\bar\chi) + O(\chi^2/m_0^2)
\eeq
we see that a change in renormalization scale can be absorbed
by shifting the $L_i$. We have checked (for $N=3$)
that the scale dependence of the $L_i$ in QCD does render
$M^2_{AB}$ independent of the renormalization scale. 

\makefig{fpi}{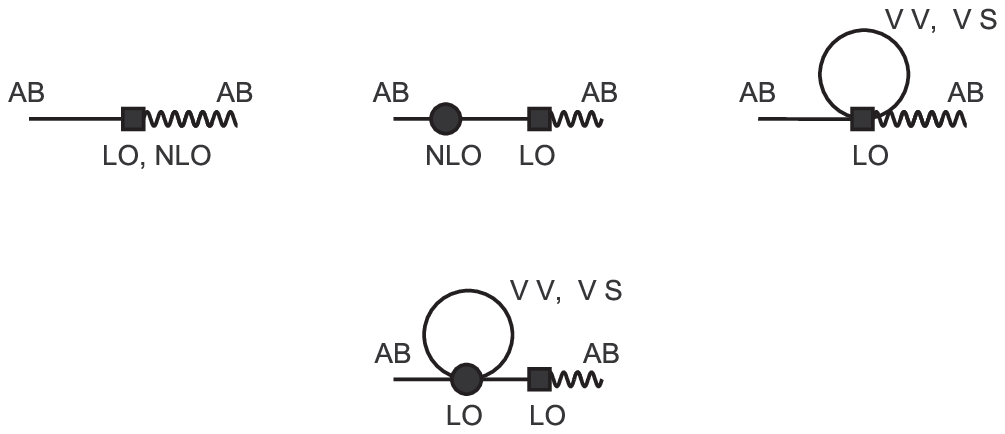}{2.2}{diagrams contributing to $f_{AB}$. The wavy
  line represents the insertion of the axial current operator
  $j_{5AB}^\mu(p)$. ``VS'' stands for Valence-Sea mesons with $V=A,B$
  and $S=1,2,3$.}

To determine the meson decay constant, $f_{AB}$, the axial current 
$j_{5AB}^{\mu}$ is calculated to NLO, and then we evaluate
the matrix element
\beq
\left< 0|j_{5AB}^\mu|\pi_{AB}(p)\right> =i\sqrt{2} f_{AB}\, p^\mu,
\eeq
using the diagrams of Fig.~\ref{fpi}.
The result is 
\beq
f_{AB} &=& f \left(1 + \delta^f_{\rm tree} + \delta^f_{\rm VS\ loop} 
+ \delta^f_{\rm VV\ loop} \right) 
\label{eq:fpiNLO}\\
\delta^f_{\rm tree}&=& \frac{4N}{f^2} \bar\chi L_4
+\frac{2}{f^2}(\chi_A+\chi_B) L_5 
\label{eq:dfpitree}\\
\delta^f_{\rm VS\ loop}&=& 
- \sum_{i=1,3} N_i \frac{1}{16 \pi^2 f^2}\frac{\chi_A+\chi_i}{8}
\log\left({\chi_A+\chi_i \over 2}\right)  \ + \ \(A\leftrightarrow B\) 
\label{eq:dfpiVSloops}\\
\delta^f_{\rm VV\ loop}&=& \frac{1}{4N}\frac{1}{16 \pi^2 f^2}
\left\{ - D_A -D_B \right.
\nonumber\\
&&+ {\log(\chi_A/\chi_B) \over (\chi_A-\chi_B)}
\left[\chi_A D_A + \chi_B D_B + (\chi_A-\chi_B)^2\right] 
\label{eq:dfpiVVloops}\\
&&\left. + \left(\chi_\pi R_\pi (\chi_B\!-\!\chi_A)
\left[{\log(\chi_\pi/\chi_A) \over \chi_A-\chi_\pi} -
{\log(\chi_\pi/\chi_B) \over \chi_B-\chi_\pi}\right] 
+\(\pi\leftrightarrow \eta\) \right) \right\}
\nonumber
\eeq
where 
\beq
D_A = {\prod_{i=1,3} (\chi_i-\chi_A) \over (\chi_\pi-\chi_A)(\chi_\eta-\chi_A)}
\,, \label{eq:DA}
\eeq
and $D_B$ obtained by $\(A\leftrightarrow B\)$,
are the coefficients of the double poles in the
neutral propagators.
It is straightforward to see that the scale dependence can be absorbed
by shifts in $L_4$ and $L_5$, and we have checked that these shifts
are consistent with standard results for $N=3$.

As noted in the introduction, our formulae can be used to extract
the GL coefficients by fitting to results from simulations.
PQ simulations allow one to
vary the valence and sea-quark masses independently, 
and thus to separately determine $L_{4-6}$ and $L_8$.
In fact, since the NLO analytic terms, 
Eqs.~(\ref{eq:mpiNLO}) and (\ref{eq:dfpitree}),
depend on the quark masses only through the combinations
$\chi_A+\chi_B$ and $\bar\chi$,
one need only consider degenerate sea quarks.\footnote{%
Note that the NLO analytic dependence of $M_{AB}^2$ 
is {\em not} the most general quadratic term, 
symmetric under $A\leftrightarrow B$, and vanishing
when $\chi_A=\chi_B=0$. Such a form would contain a 
term proportional to $(\chi_A-\chi_B)^2$, 
which is in fact forbidden by chiral symmetry.}
Indeed, to extract $2 L_8-L_5$, 
the combination which determines whether $m_u=0$,
it is sufficient to use a single sea quark mass
(as long as it is light enough that the formulae apply),
and vary the valence quark masses.
By contrast, with unquenched simulations, 
one would have to use non-degenerate
sea quarks to separately determine all four $L$'s.

The most likely practical applications of these results are for
simulations done with 2 rather than 3 types of nondegenerate sea quarks.
For QCD this would correspond to the limit of exact isospin symmetry.
The results for this case can be obtained from those given above
by carefully taking the limit $\chi_2\to\chi_1$.
When two types of quark are degenerate, one of the neutral
sea-quark eigenstates becomes an exact flavor non-singlet,
and we choose this to be the pion.
Thus we have $\chi_\pi=\chi_1=\chi_2$.
The remaining neutral state has mass 
$\chi_\eta= \chi_1+\chi_3-\bar\chi$.
In this limit, the residues simplify, e.g.
\beq
R_A &\longrightarrow& 
\frac{\(\chi_A-\chi_1\)\(\chi_A-\chi_3\)}
{\(\chi_A-\chi_B\)\(\chi_A-\chi_{\eta}\)}\\
R_\pi &\longrightarrow& 0 \\
R_\eta &\longrightarrow&
 \frac{\(\chi_\eta-\chi_1\)\(\chi_\eta-\chi_3\)}
{\(\chi_\eta-\chi_A\)\(\chi_\eta-\chi_B\)}
\eeq
with a similar cancelation in the $D_{A,B}$.
With these changes, the formulae given above still hold.

We have checked our results in two ways. First, if we take all
sea quarks to be degenerate, we obtain the
results of Ref.~\cite{SSPQ}.\footnote{Except for the
following typos in Ref.~\protect\cite{SSPQ}, pointed out 
by Jochen Heitger, Rainer Sommer and Hartmut Wittig:
in Eqs.~(18), (19) and (20) $\alpha_4$ should be replaced by $\alpha_4/2$.}
Second, we can consider the unquenched limit. 
By choosing $\chi_A$ and $\chi_B$ to be equal
to combinations of $\chi_{1-3}$ we obtain the correct one-loop form for the
masses of the $\pi^+$, $K^+$ and $K^0$. 

The residues $R_x$ and $D_x$ are singular when pairs of the
$\chi$'s become degenerate, e.g. $\chi_A\to\chi_B$ and $\chi_A\to\chi_\pi$.
As expected, however, these singularities cancel 
in the full expressions for $M_{AB}$ and $f_{AB}$,
which are analytic functions of the quark masses except
in the massless limit.

In Ref.~\cite{SSPQ}, it was emphasized that the one-loop corrections
can diverge when the
valence-quark masses are sent to zero at fixed sea-quark mass,
leading to a breakdown of chiral perturbation theory.
This discussion was based on the results for degenerate sea quarks,
and we can now generalize it to non-degenerate sea-quarks.
For the meson masses, the possibly divergent contribution is
$\delta^M_{\rm loop}$, and we see that this only diverges if
both $m_A$ and $m_B$ vanish, in fixed ratio, but not if only one
vanishes.  For the decay constant the pattern is opposite:
$\delta^f_{\rm VV\ loop}$ diverges if one of the valence masses vanishes,
but not if both do in fixed ratio.
This is the same pattern of divergences as for degenerate sea quarks;
the sea quark masses only influence the coefficients of the 
divergent $\log\chi_A$ and $\log\chi_B$ terms.
In the following section we discuss the practical implications
of these divergences.

It was also noted in Ref.~\cite{SSPQ} that one can form
combinations of the squared meson masses and decay constants from which
the analytic correction terms ($\delta_{\rm tree}$) cancel.
One thus predicts these combinations
in terms of the quark masses and the leading order
chiral coefficients $\mu$ and $f$,
up to NNLO corrections.
Studying these combinations in simulations 
allows one to test the applicability
of NLO chiral perturbation theory.
What we want to point out here is that exactly
the same quantities can be used for non-degenerate sea-quarks:
the $L_i$ still cancel.
We do not, however, give the explicit expressions since they are 
lengthy and unilluminating.

Finally, we discuss to what extent our formulae can be applied to
lattice results obtained at non-zero lattice spacing.
For definiteness, we first consider a calculation using Wilson fermions,
in which we work at a fixed bare coupling and vary the bare valence
and sea quark masses. Meson properties are of course calculated
in lattice units, i.e. we obtain $M_\pi a$ and $f_\pi a$.

There are two effects of working at a non-zero lattice spacing.
The first is that chiral symmetry, upon which our calculation is based,
is broken explicitly.
This symmetry breaking can, however, 
be incorporated into the chiral Lagrangian framework,
as shown in Ref.~\cite{ShSi}.
The result is that all effects of the explicit symmetry breaking
are of $O(a)$, except for the additive renormalization of the quark masses.
Because of this, our formulae are valid, up to corrections of $O(a)$,
as long as one uses so-called vector or axial Ward identity quark masses.

Note that the corrections of $O(a)$ cannot be incorporated into
our formulae by simply introducing $a$ dependence into the parameters of the
chiral Lagrangian parameters, $f$, $\mu$ and the $L_i$.
There are additional unknown constants which enter.

The second effect is that the lattice spacing itself depends on
the quark masses, at fixed bare coupling.
This introduces an additional mass 
dependence into quantities expressed in lattice units.
We note, however, that the mass dependence of $a$ is a discretization
effect of $O(am)$ induced by the 
explicit chiral symmetry breaking~\cite{ALPHA}. 
Indeed, for non-perturbatively on-shell-improved Wilson fermions
one can calculate this dependence, with the result
\begin{equation}
a(m) = a(m\!=\!0) [1 - c\, a \overline m + \dots ]\,, \qquad
c = {g\ b_g(g) \over 2 \beta_{\rm LAT}(g)}\,,
\end{equation}
where $\overline m$ is the average sea-quark mass,
$b_g(g) \propto g^2 + O(g^4)$ is an improvement coefficient
introduced in Ref.~\cite{ALPHA}, and $\beta_{\rm LAT} = - d g/ d\ln a
\propto g^3 + O(g^5)$ is the lattice $\beta$-function.
Note that this $O(a)$ effect can be absorbed by shifting the
GL coefficients as follows:
\begin{equation}
L_{4,6} \rightarrow L_{4,6} - {c a f^2 \over 8 N \mu} \,.
\end{equation}
Alternatively, one could adjust the
bare coupling as the quark masses are varied
so as to keep the lattice spacing fixed.

In summary, our formulae are approximately valid 
for meson properties expressed in lattice units, with the
errors being of $O(a)$. Some, but not all, of these discretization
errors can be absorbed into $a$ dependence of the parameters of
the chiral Lagrangian. With staggered or overlap fermions the
errors would instead be of $O(a^2)$.
Since discretization errors can still be substantial at present
lattice spacings, it may be better to
extrapolate first to $a=0$, and then fit to the predicted forms.

\section{Behavior of the Chiral Expansion}
\label{sec:anal}

In the framework of chiral perturbation theory (regardless of quenching) 
one assumes that for any quantity calculated to a given order in the 
chiral expansion, higher order terms are smaller. Once the unknown 
couplings and parameters of the theory are determined 
(e.g. by experiment, or by fitting to lattice data) the consistency 
of the computation can be checked numerically. 
This is the main purpose of the current section.

We can make this check because, using experimental data,
we have reasonable estimates for the actual values of 
some of the Gasser-Leutwyler coefficients. Thus we can use our results
to predict the meson masses and decay constants  in PQ simulations.
Of course, these predictions are approximate because we only
know the $L_i$ approximately---this is where the 
lattice results themselves come in---but
they give us a reasonable idea of how the chiral expansion behaves.

We consider QCD with exact isospin symmetry, i.e.
$N_1=N_2=N_3=1$ and $m_1=m_2=m_u$ and $m_3=m_s$.
The meson masses and decay constants then depend on the
seven parameters $f$, $\chi_u$, $\chi_s$, $L_{4-6}$ and $L_8$.
We want to choose these parameters so that the four charged meson
quantities $M_\pi$, $M_K$, $f_\pi$ and $f_K$ take their
experimental values. In order to match the number of
parameters and observables, we take as starting values the
GL coefficients quoted in Ref.~\cite{DGH}.
These are based on experimental results
($L_5(M_\eta)\approx 2.3\ 10^{-3},\ L_8(M_\eta)\approx 1.2\ 10^{-3}$) 
and the large $N_c$ limit ($L_4(M_\eta)\approx L_6(M_\eta)\approx 0$).
We then take our four free parameters to be $f$, $\chi_u$, $\chi_s$
and the scale, $\Lambda_L$, at which $L_i$ take the values just quoted. 
In effect, this moves us through the space of $L_i$
on a particular path, which, when $\Lambda_L \approx M_\eta$,
is consistent with our knowledge about the $L_i$. 
We claim no fundamental basis for this path---we use it for simplicity.
It allows us to determine the dimensionless quantities
$f/\Lambda_L$, $\chi_u/f^2$ and $\chi_s/f^2$
by fitting the ratios $M_\pi/f_K,\ M_K/f_K$ and $f_K/f_\pi$,
and then to determine $\Lambda_L$ by requiring, say, $M_\pi=140\;$MeV.  
We find $\Lambda_L=M_\eta(1-.0021)$ (using $M_\eta=547\;$MeV),
so that the fitted $L_i$'s are, in fact, close to the inputs.
The other outputs are $f=85$MeV, 
 $\chi_u/\chi_s=0.044$ and $\chi_s=(673\mbox{MeV})^2$.

We stress that we are not claiming that we have found a unique set
of parameters. There is a region in the space of the $L_i$ which
can describe the experimental observables, and for which 
it turns out that the chiral
expansion is under reasonable control. We have picked, somewhat
arbitrarily, one point in this region.

We can now explore the behavior of the chiral expansion 
as a function of the four quantities 
\beq
\left({\chi_A \over \chi_s},\,{\chi_B \over \chi_s},\,
{\chi_1 \over \chi_s},\, {\chi_3 \over \chi_s}\right) \,.
\eeq
We have chosen to normalize the various quark masses relative to
the physical strange quark mass, so that a ratio of unity
represents the outer limit of where one would expect chiral perturbation
theory to be reliable.
We consider two types of two-dimensional cross-section of the
parameter space: $(y,y,x,1)$ and $(y,1,x,1)$. 
In both cases $y$ corresponds to a valence mass
while $x$ is proportional to a sea quark mass. 
We name these cross-sections the ``$\pi$-plane'' and the 
``$K$-plane'' because the unquenched line $y=x$ in the former
describes a pion-like meson made up of two identical light quarks, 
whereas, in the latter, $y=x$ corresponds to a kaon-like meson. 

We examine the relative size of the NLO contributions by plotting
\beq
(M_{AB}^2)_{\mbox{NLO}} /(M_{AB}^2)_{\mbox{LO}}=
(\delta^M_{\rm tree} + \delta^M_{\rm loop})\equiv \delta^M 
\eeq
and
\beq 
(f_{AB})_{\mbox{NLO}}/(f_{AB})_{\mbox{LO}}=
\delta^f_{\rm tree} + \delta^f_{\rm VS\ loop} 
+ \delta^f_{\rm VV\ loop}\equiv \delta^f 
\eeq 
[see Eqs.~(\ref{eq:MABNLO}), (\ref{eq:fpiNLO})].
In each plane, these functions are plotted along rays
emerging from the origin at angles 
$15^\circ,\ 30^\circ,\ 45^\circ,\ 60^\circ$ and $75^\circ$
with respect to the $x$-axis, 
confined to the unit square (Fig.~\ref{thetas}).
The $45^\circ$ line corresponds to unquenched theories,
the other lines to PQ theories.
The plots are shown in Figs.~\ref{fpirays}-\ref{MKrays}.
We also show a contour plot of
$\delta^f$ in the K-plane in Fig.~\ref{ContourAll}.

\makefig{thetas}{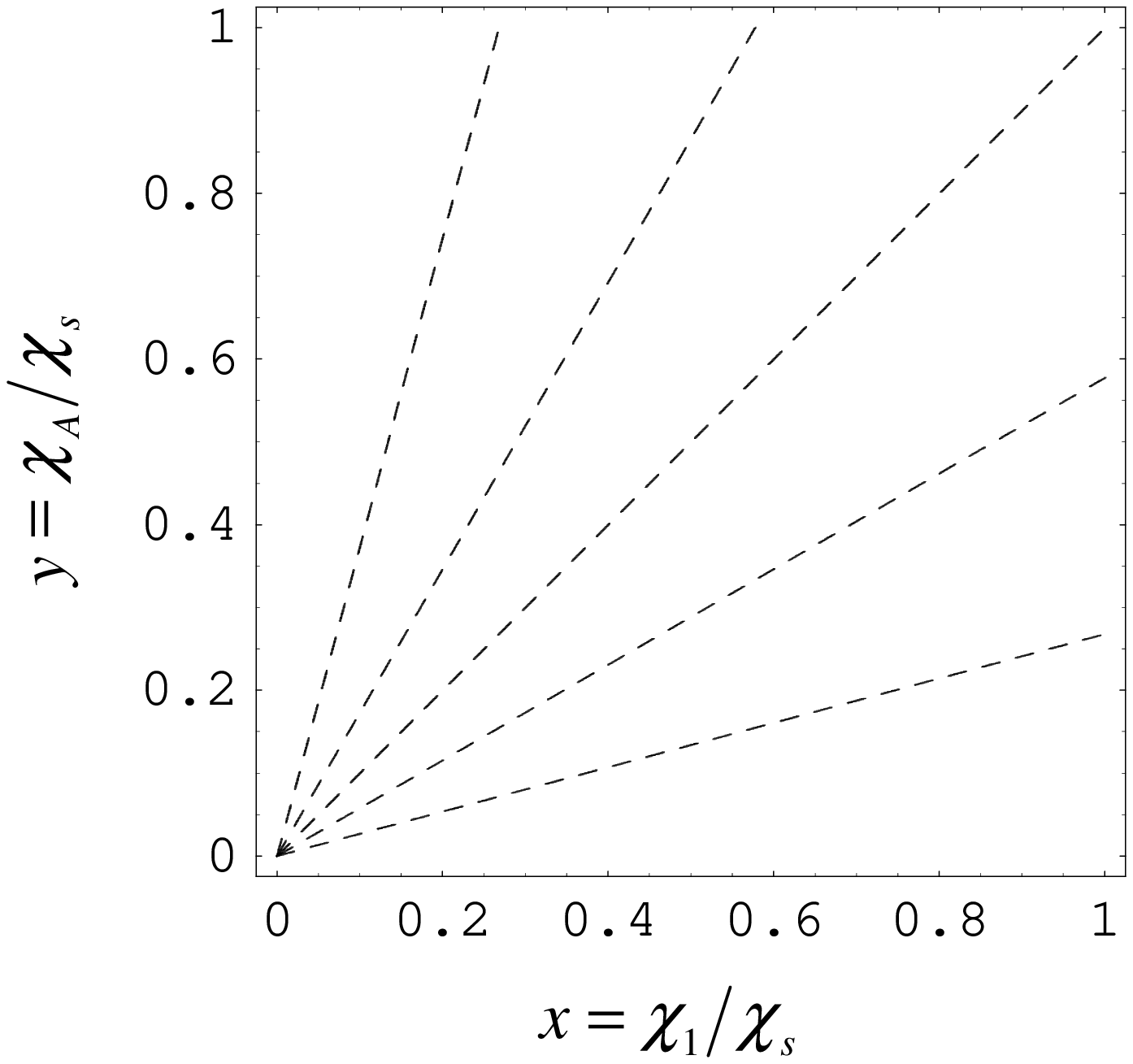}{3}{The dashed lines 
are rays in the $\pi$-plane or $K$-plane along which 
$\delta^f$ and $\delta^M$ are plotted in Figs.~\ref{fpirays}-\ref{MKrays}.}
\makefig{fpirays}{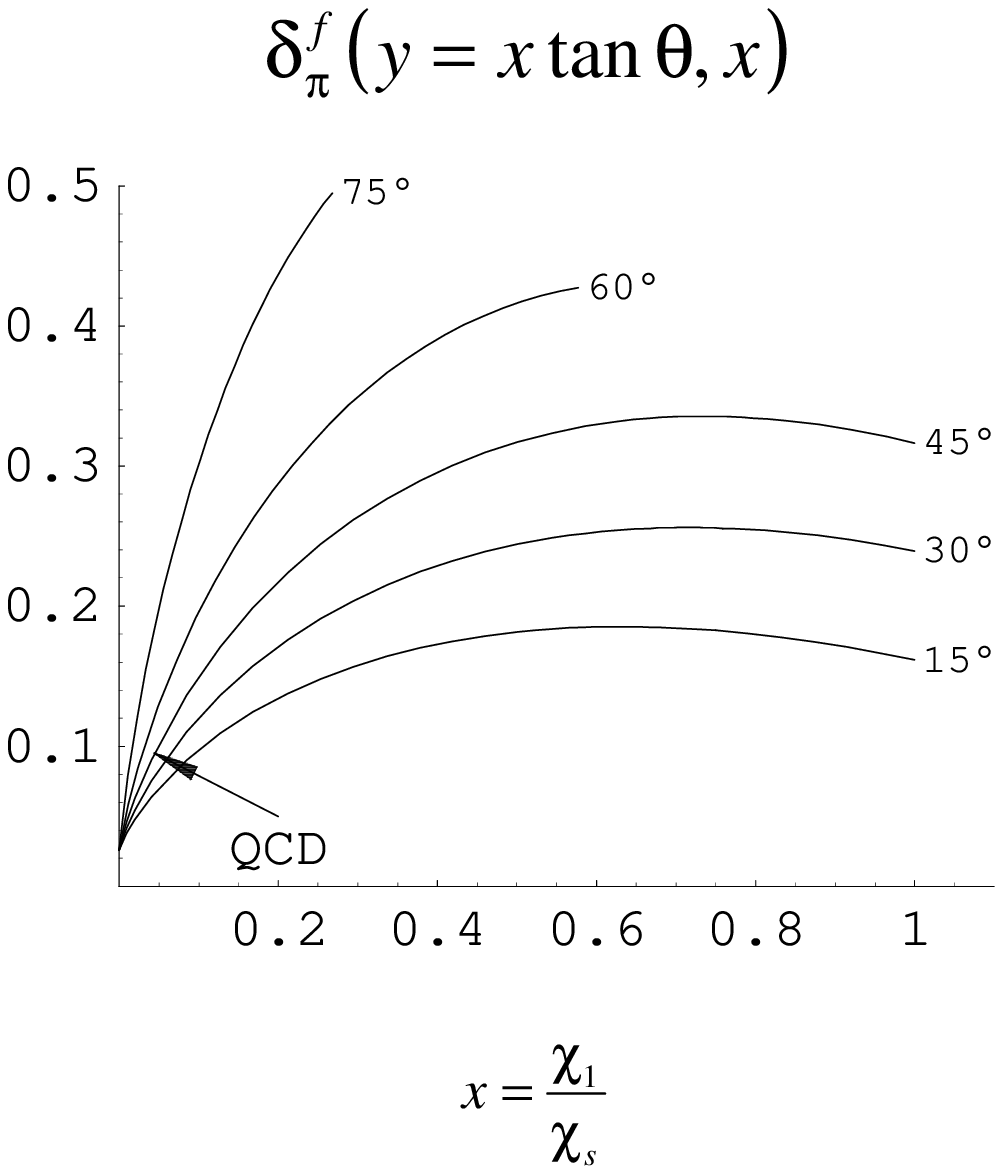}{3.5}{$\delta^f$ in the $\pi$-plane, 
$(y,y,x,1)$, is plotted along rays of angle $\theta$ with respect 
to the $x$-axis (see Fig.~\ref{thetas}). The values of theta in 
degrees are indicated next to the corresponding curves. 
The point that corresponds to the ``physical'' pion, 
$(\chi _u ,\chi _u ,\chi _u ,\chi _s )$ is labeled ``QCD''.}
\makefig{fKrays}{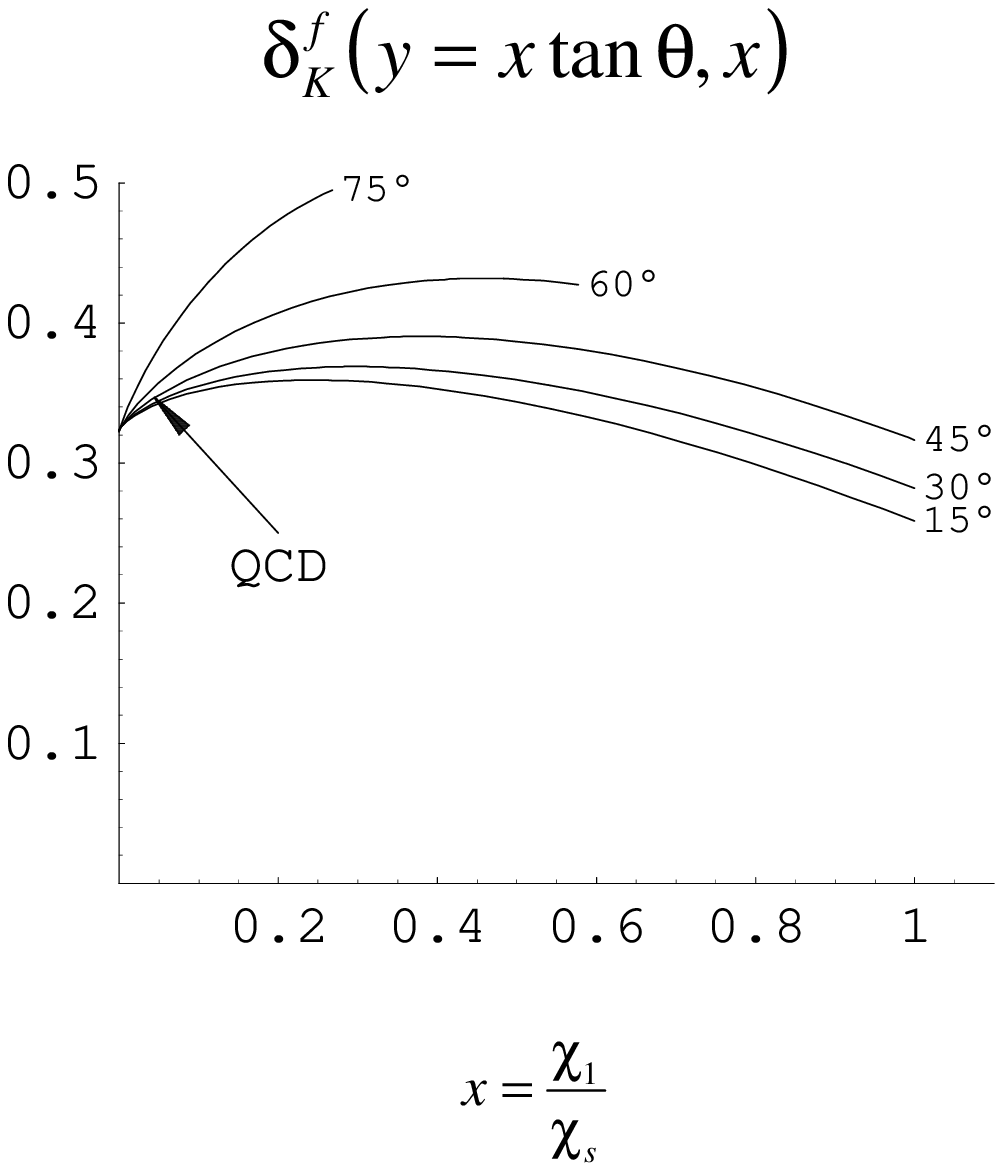}{3.5}{Plots of $\delta^f$ in the $K$-plane.}
\makefig{Mpirays}{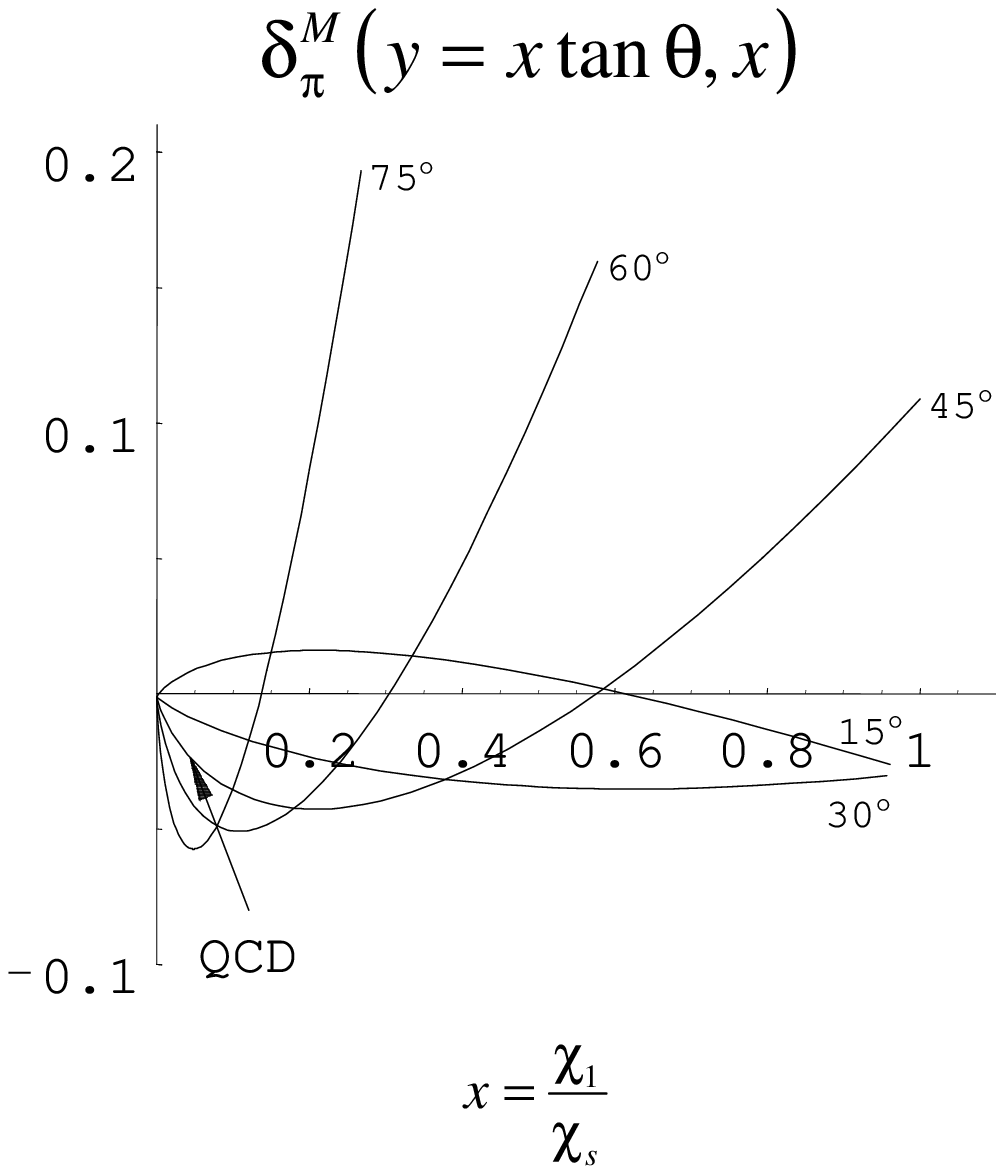}{3.5}{Plots of $\delta^M$ in the $\pi$-plane.}
\makefig{MKrays}{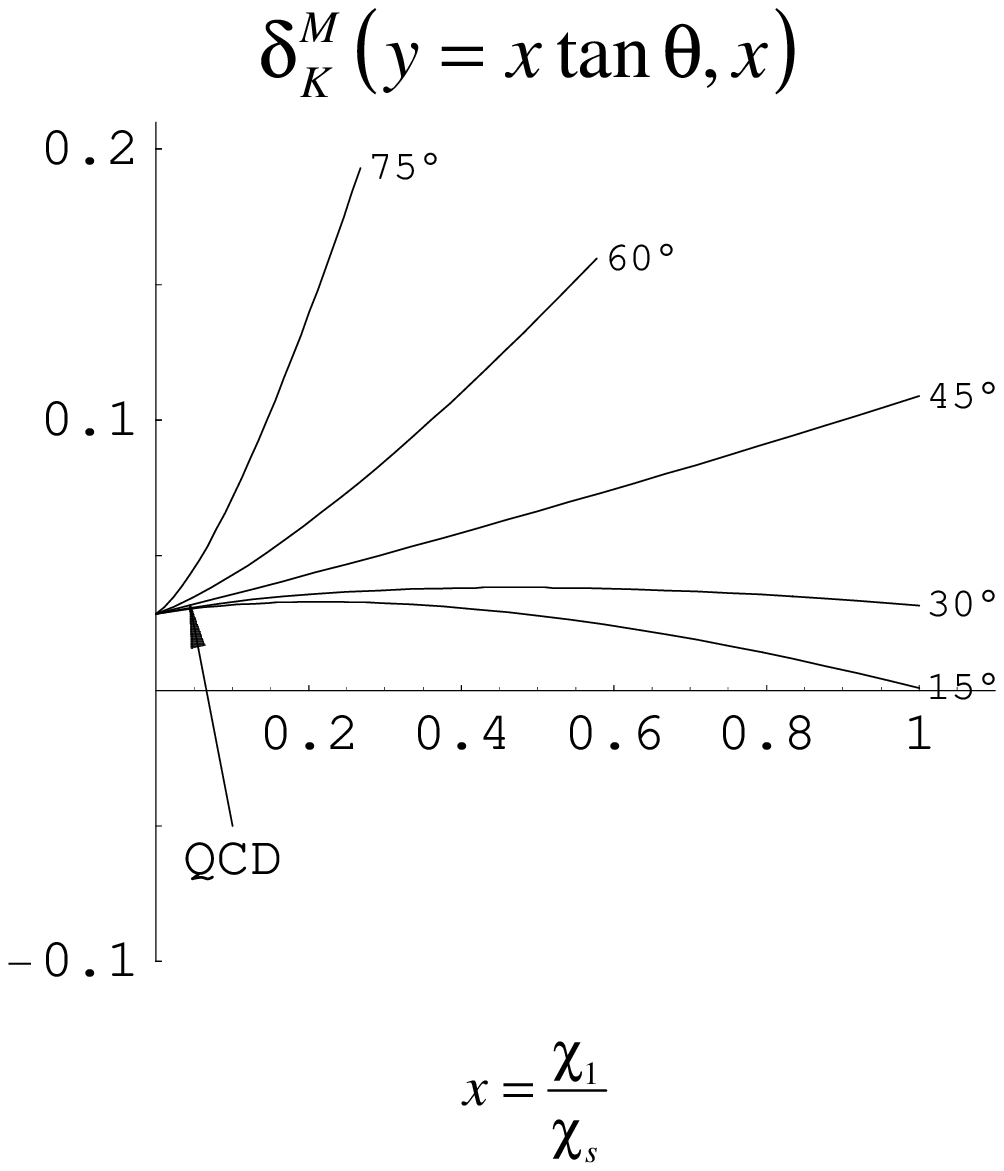}{3.5}{Plots of $\delta^M$ in the $K$-plane.}

\makefig{ContourAll}{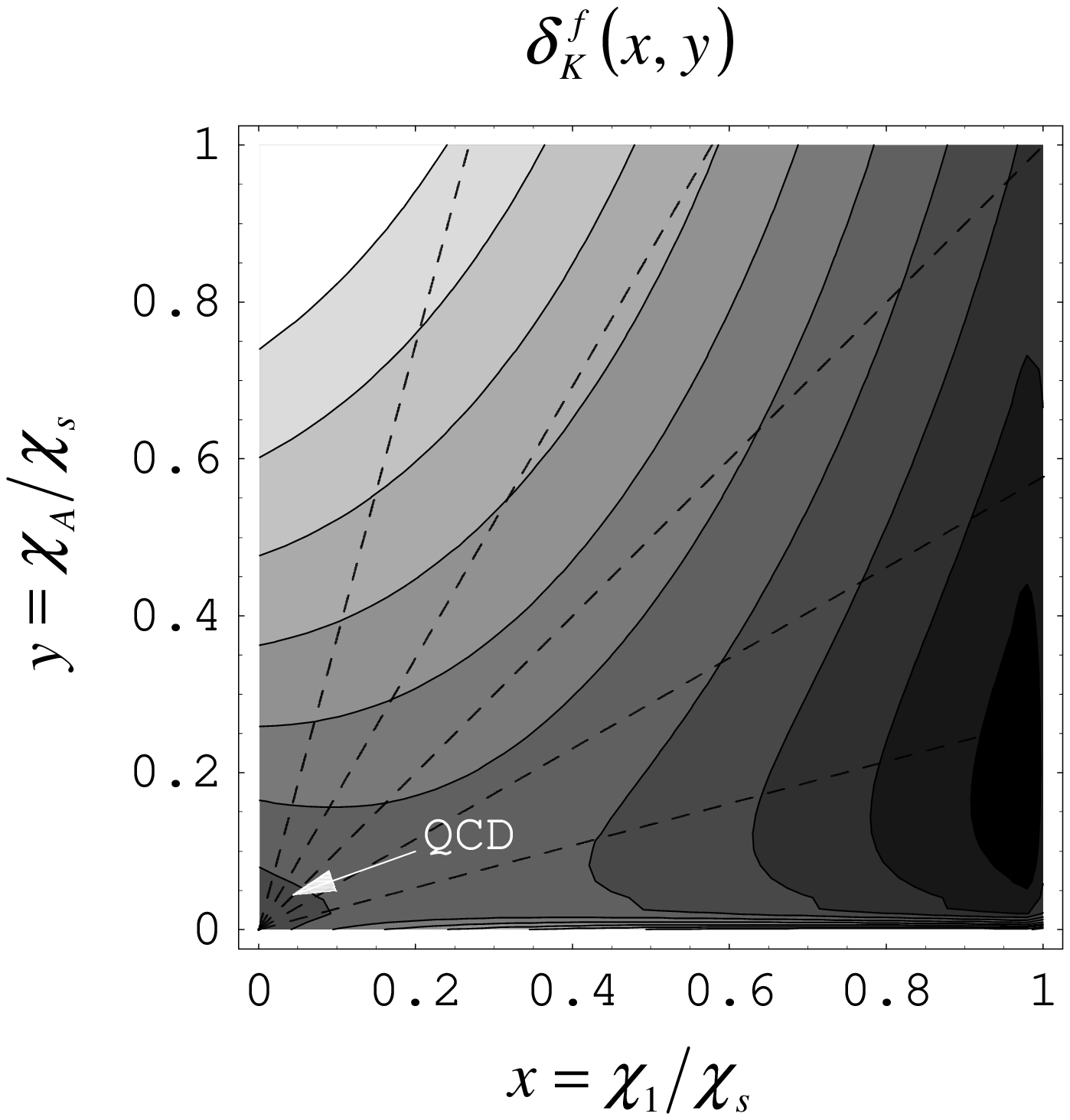}{4}{Contours of $\delta^f_K$. 
The values of $\delta^f_K$ range from $<.275$ in the black region 
on the right up to values $>.5$ in the white region. There is a 
difference of .025 between the values of $\delta^f_K$ on adjacent contours}

The first conclusion that can be drawn from these figures is that chiral
perturbation theory for QCD to one-loop order is reasonably convergent
throughout the PQ region defined by $\chi_x<\chi_s$, $x=A,B,1,3$.
With our parameters, the least convergent quantity is $f_K$.
We also note that, generally speaking,
the expansion is better behaved when one increases
the masses along the lines at small angles. In other words, the
expansion may be more reliable when valence masses are smaller
than sea-quark masses, and {\em vica-versa}.
This is good news for simulations, since pushing to small valence
quark masses is relatively cheap.

Of course, as noted in the previous section, the valence quark
masses cannot be pushed too low at fixed sea masses. 
At some point the corrections
$\delta^{M}$ (in the $\pi$-plane) and $\delta^f$ (in the K-plane) diverge.
This is not apparent, however, from Fig.~\ref{ContourAll}.
To observe the divergence we show in Fig.~\ref{ContourLogs} the
region close to the $x$-axis, which does reveal the expected features.
Though of theoretical interest, it seems that the breakdown of the 
chiral expansion due to enhanced chiral logarithms occurs 
only in a tiny region of parameter space and is therefore of little 
or no practical significance. 

Our final comment concerns the importance of including the
non-analytic terms in fits to PQ data.
The tree level contributions alone would have 
produced  only straight lines in Figs.~\ref{fpirays}-\ref{MKrays}. The
curvature seen in the graphs is due to the logarithms
originating in loop diagrams. An attempt to model data collected in
the heavy quark mass region with only the tree level terms will
clearly lead to a significant systematic errors in the extrapolations
to QCD values. 

\makefig{ContourLogs}{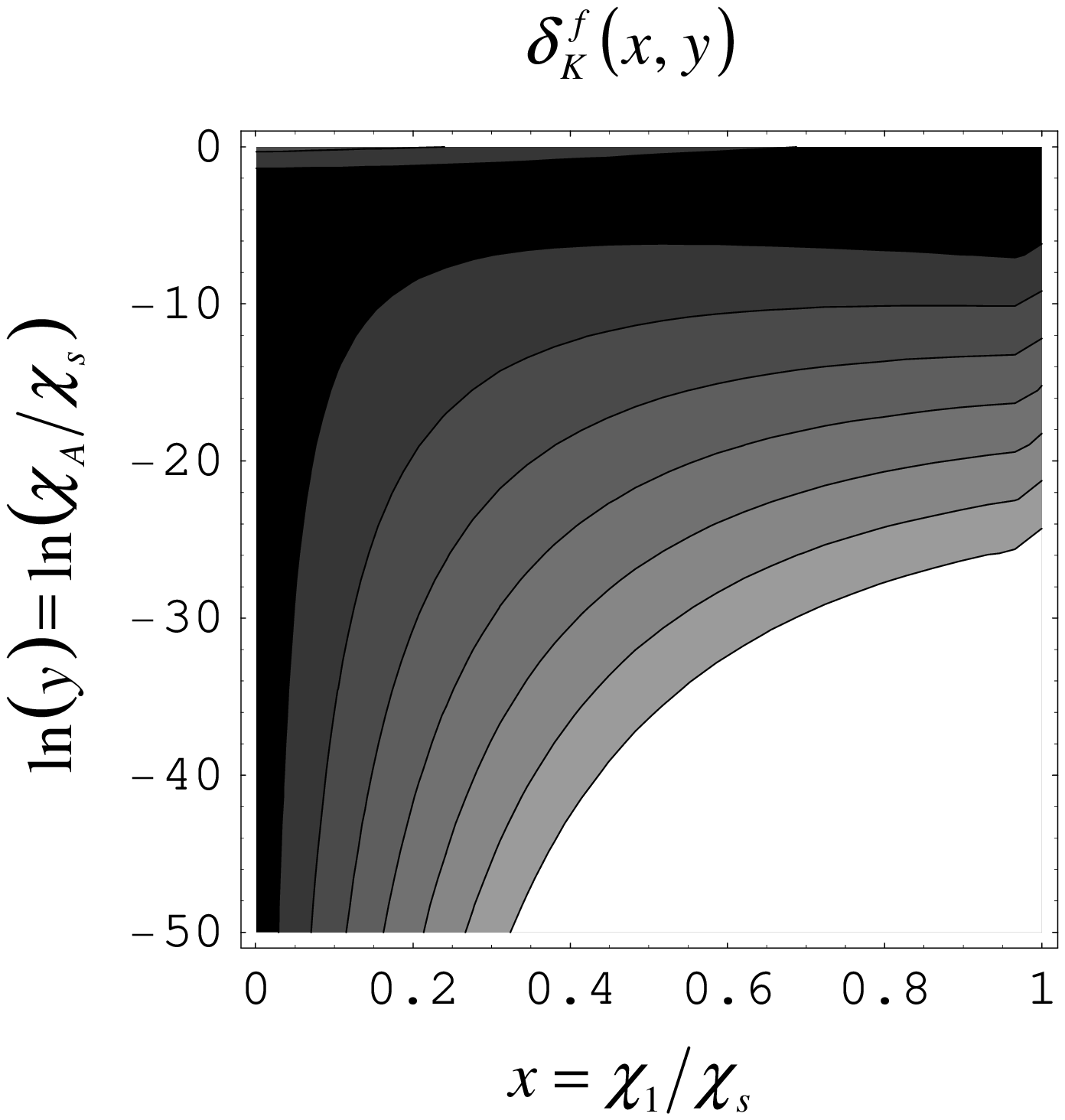}{4}{Breakdown of the chiral 
expansion for small values of valence quark mass and a fixed sea 
quark mass. In the white region, the expansion breaks down as the 
NLO term in the expression for $f_K$ becomes greater than the LO 
term ($\delta^f_K>1$). The following contours mark changes of .1 
in $\delta^f_K$, so that the darkest region corresponds to $\delta^f_K<.4$}

\section{Determining $L_7$}
\label{sec:L7}

As noted above, the GL coefficients $L_4$, $L_5$, $L_6$ and $L_8$ can
be obtained using PQ simulations with degenerate sea quarks.
Simulations with non-degenerate sea quarks are useful but not essential.
In this section we consider the other coefficient which enters
into the NLO expressions for the physical meson masses, namely $L_7$.
This contributes only to flavor-diagonal mesons, i.e. the $\eta$ and 
$\pi_0$ in QCD. Since the contribution is proportional to
quark mass differences, the largest effect is on the ${\eta}$ mass.
This is conveniently isolated using the violation of the mesonic
Gell Mann--Okubo relation~\cite{DGH}
\beq
4 m_K^2 -m_\pi^2 - 3 m_\eta^2
&=& {8 \over f^2} {4 (m_K^2 - m_\pi^2)^2 \over 3}
(L_5-6 L_8 - 12 L_7) \nonumber\\
&&\mbox{}+ {\left[3 m_\eta^4 \log(m_\eta^2) -m_\pi^4 \log(m_\pi^2)
- 4m_K^4 \log(m_K^2)\right] \over 16 \pi^2 f^2}\,.
\label{eq:meta}
\eeq
Here we have given the result in the isospin-symmetric limit,
i.e. with two degenerate quarks of mass $\overline m = (m_u + m_d)/2$,
and a single strange quark.\footnote{Note that the one-loop expressions
for the dependence of $m_K$ and $m_\pi$ on quark masses can be
obtained from the general results above by setting $N_1=N_2=N_3=1$, $m_2=m_1$,
and choosing the valence masses to be equal to the appropriate sea quark
masses.}

Thus one method for determining $L_7$ is 
to calculate $m_\eta$, $m_K$ and $m_\pi$ 
in unquenched simulations with non-degenerate sea quarks and
fit to the result in Eq.~(\ref{eq:meta}). 
In this section we show how one can obtain
$L_7$ using degenerate sea quarks by
taking advantage of PQ simulations.

A clue on how to proceed is provided by the fact that the
${\eta}$ propagator contains contractions
in which the quark propagators are disconnected (``hairpin'' contractions).
Close to the degenerate limit,
these contractions are proportional to $(m_s-\overline m)^2$,
due to cancelations between light and strange quark propagators.
Comparing to Eq.~(\ref{eq:meta}), 
and noting that $m_K^2-m_\pi^2 \propto m_s-\overline m$,
it is plausible that the $L_7$ contribution
is related to the disconnected contraction, and that by studying this
contraction in the PQ theory one can determine $L_7$.
This is indeed what we find.

First, we recall the general form of the tree level propagators.
In section~\ref{sec:calc} we showed that
the propagators for flavor off-diagonal mesons, $G_{ab}^C$, 
have an ordinary single pole at the meson mass [Eq.~(\ref{eq:GC})], 
while $G_{ab}^N$, the ``neutral'' or flavor-diagonal propagators, 
have a double pole at the same mass 
if $a=b$ or $m_a=m_b$ [Eq.~(\ref{eq:GNtree})].
As we show below, these general forms prevail also at NLO:
\beq
G_{AB}^C(p)\big|_{m_A=m_B}
&=& {{\cal Z_A} \over p^2 + M_{AA}^2}
+ \mbox{non-pole} \,,\label{eq:DPQ1} \\
G_{AB}^N(p)\big|_{m_A=m_B}
&=& {{\cal Z_A} {\cal D} \over (p^2 + M_{AA}^2)^2} 
+ \mbox{single-pole}+\mbox{non-pole} \,,\label{eq:DPQ2} \\
G_{AA}^N(p) &=& G_{AB}^N(p)\big|_{m_A=m_B}+G_{AB}^C\big|_{m_A=m_B} 
\nonumber \\
&=&  {{\cal Z_A} {\cal D} \over (p^2 + M_{AA}^2)^2} 
+ \mbox{single-pole}+\mbox{non-pole} \,.
\label{eq:DPQ3}
\eeq
Here $M_{AA}$ is a shorthand for $M_{AB}(m_A=m_B)$,
and its NLO expression is given in Eq.~(\ref{eq:MABNLO}).
Note that these equations are valid only if $A$ and $B$ are
partially quenched rather than unquenched (i.e. $m_A\ne m_i$), 
and apply only close to the poles at $p^2= M_{AA}^2$.
In particular, the neutral propagators have additional poles,
at the masses of the neutral sea-quark mesons.
These are not of interest here.

The quantity we propose to use to determine $L_7$ is ${\cal D}$, 
the (suitably normalized)
coefficient of the double pole in the neutral propagators.
An equivalent definition in terms of lattice observables is 
\beq
{\int d^3 x \langle \pi_{AA}(t, \vec x) \pi_{BB}(0) \rangle \over
 \int d^3 x \langle \pi_{AB}(t, \vec x) \pi_{BA}(0) \rangle}\bigg|_{m_A=m_B}
{\longrightarrow \atop {\rm t\to\infty} }\ 
{{\cal D} t \over 2 M_{AA}} \,.
\label{eq:Dlat}
\eeq
Here the flavor indices are chosen to select the
``disconnected'' (numerator) and ``connected'' (denominator) 
contractions contributing to the $\pi_{AA}$ propagator.\footnote{
This choice simplifies the numerical calculation,
but is not necessary. It follows from
Eq.~(\protect\ref{eq:DPQ3}) that the numerator could be replaced by
$\langle \pi_{AA}(t, \vec p=0) \pi_{AA}(0) \rangle$, i.e. the
temporal Fourier transform of $G_{AA}^N$.
This changes the $t-$independent part of the ratio but leaves
${\cal D}$ unchanged.}
At large times, the denominator is dominated by the single-pole
contribution of the pseudo-Goldstone boson of mass $M_{AA}$. 
Only double-pole contributions to the numerator lead to the
ratio growing linearly with $t$, and ${\cal D}$ measures their size.
We note also that our ratio is the standard one 
used in studies of artifacts in the quenched theory.

We claim that ${\cal D}$ is a ``physical'' quantity in the PQ theory,
on a par with the valence meson masses such as $M_{AA}$.
In support of this claim, we note that ${\cal D}$ is
determined from the long distance properties of correlators, 
and is independent of the choice of interpolating fields
(because it is determined by a ratio).
In particular, we claim that ${\cal D}$
can be calculated using the effective low-energy theory 
with the same level of  reliability as the meson masses.\footnote{Proving 
this claim rigorously seems difficult, given the lack of a 
Hamiltonian formulation of the PQ theory. Our point, however, is
that the calculation of ${\cal D}$ is as well controlled as that
of meson masses.}
We stress that the coefficients of the single poles, unlike the
double pole, do depend on the choice of interpolating fields and are
not quantities which can be predicted using chiral perturbation theory.

In PQChPT, we obtain $\mcal{D}$ 
by calculating the propagators $G^C_{ab}$ and $G^N_{ab}$ at NLO
and using Eqs.~(\ref{eq:DPQ1})-(\ref{eq:DPQ3}). 
The tree level result for ${\cal D}$ 
can be read off from Eqs.~(\ref{eq:GC}) and (\ref{eq:GNtree})
\beq
\label{eq:DLO}
{\cal D} =  - {1\over N} 
{(\chi_{1} - \chi_{A}) (\chi_{2}-\chi_{A}) (\chi_{3} - \chi_{A})\over
 (\chi_{\pi} - \chi_{A}) (\chi_{\eta} - \chi_{A}) }
\eeq
Note that the residue of the double-pole vanishes whenever the
valence quark mass equals any of the sea quark masses.
This must be the case since one is then considering a correlator
which could be constructed entirely from sea quarks, and thus is
physical, and cannot contain double poles.

The dependence of ${\cal D}$ on $L_7$ begins at one-loop order.
We have calculated ${\cal D}$ to this order 
only for the case of degenerate sea quarks ($m_1=m_2=m_3$).
Details are given in app.~\ref{app:DNLO}. The result is
\beq
{\cal D} &=& -{1\over N} (M_{SS}^2-M_{AA}^2) 
-{16\over f^2} \left(L'_7 - {f^2 \over 16 N m_0^2} +{L_5 \over 2N}\right) 
(M_{SS}^2-M_{AA}^2)^2 
\nonumber\\ 
&&\mbox{} 
+ {1\over 16 \pi^2 f^2} \left( {1\over2}[\chi_S-\chi_A]^2\log\chi_{SA}
+ \chi_A^2 \log(\chi_A/\chi_{SA}) + 
\chi_S^2 \log(\chi_S/\chi_{SA}) \right) \,,
\label{eq:DNLO}
\eeq
where we use $S$ to denote the sea-quark, 
and $\chi_{SA}=(\chi_S+\chi_A)/2$.
The first term is the same as the tree-level
result with one-loop corrected meson mass-squareds replacing quark masses.
The second term is the analytic term containing the $L'_7$ dependence.
As advertised, $L'_7$ and $m_0^2$ appear in the appropriate combination to
be combined into the standard $L_7$ [Eq.~(\ref{eq:L7shift})]. 
The logarithmic terms, 
from wavefunction renormalization and from loop diagrams, combine
into a fairly simple form. One check on the result is that 
the anomalous dimension of $L_5$ is such that it cancels the
dependence on the choice of the scale in the logarithm
(for $N=3$, where the anomalous dimension is known).
Another is that it vanishes when $m_1=m_A$.

Thus, from the coefficient of the double pole, one can extract the
combination $2 N L_7+ L_5$. Combined with the results of the
previous sections this allows a determination of $L_7$.

\section{Conclusions}
\label{sec:conclusions}

Partially quenched theories can play an important role in
determining physical parameters of QCD.
Our results show how one can use them to simplify the
extrapolation to QCD and the determination of the GL parameters
$L_{4-8}$. In particular, it is sufficient, though not necessary,
to use degenerate sea quarks. One must, however, use three 
sea quarks.

Our approach relies on chiral perturbation theory at next-to-leading order.
An important issue is how light the sea quarks need to be
for our formulae to be sufficiently accurate.
A conservative approach is to work down to masses where the
NLO corrections themselves are 10\% of the leading order 
result, so that the missing NNLO terms are very small~\cite{SSICHEP98}. 
This requires working down to $m_{\rm sea} \approx m_s/8$.
Another approach is to look at our figures and see what mass
range is required to observe the predicted curvature.
To do so appears to require working down to at least
$m_{\rm sea}\approx m_s/4$.
In the end, this issue can be resolved using simulations themselves,
including partially quenched simulations,
to check the reliability of the NLO predictions.

This exercise will also shed light on the question of whether
the physical strange quark mass is light enough that NLO
chiral perturbation theory is applicable for QCD.
Even if the strange quark turns out to be too heavy,
the results presented here still apply to the light-quark sector, 
if we set $N=2$.

It is often observed that linear fits are adequate for
the mass dependence of physical quantities in the
range $m_{\rm sea}= m_s/2-m_s$.
Our results make clear, however, that extending such linear
fits to the chiral limit, i.e.
leaving out the curvature due to chiral logarithms,
can lead to a significant error in the determination of
physical parameters. This is most clearly illustrated by 
Fig.~\ref{fKrays}.

All the formulae we present are for infinite volume,
and thus apply for box sizes such that $M_\pi L \gg 1$.
As pion masses decrease, this requires working in boxes of increasing size.
This should not, however, be necessary.
Chiral perturbation theory remains valid
for finite $M_\pi L$, as long as $\Lambda_{\rm QCD} L\gg 1$,
and can be used to calculate the volume dependence of physical quantities.
It would be interesting to do this for the quantities we consider here.

Another interesting extension of our calculations is to include
the effect of discretization errors within the chiral Lagrangian itself,
i.e. to use the appropriate Lagrangian for the lattice theory at 
non-zero lattice spacing. This would aid in the extrapolation to the
continuum limit.

Finally, we comment on the theoretical status of our methods.
We use a Lagrangian containing the $\eta'$
(or more precisely the ``super-$\eta'$'', $\Phi_0$),
and our calculations rely on assumptions about the size of its couplings.
One can show, however, that there is a limit in which our calculation is
equivalent to that with the $\eta'$ integrated out 
non-perturbatively~\cite{ShShNP}.
In that limit (essentially $m_0^2\to \infty$) the assumptions that
we have made are valid.
One might also be concerned about the theoretical foundation of
the whole calculation: Is it justified to use
chiral perturbation theory for the unphysical PQ theory?
We have also made some progress on this issue.
One can show that, in certain cases,
derivatives of partially quenched quantities
with respect to valence quark masses
can be exactly related to
derivatives of unquenched quantities with respect to 
sea quark masses~\cite{ShShNP}.
Our NLO results are consistent with these exact relations.

\section*{Acknowledgments}
We thank David Kaplan and Ann Nelson for useful conversations.
This work was supported in part by U.S. Department of Energy
Grant No. DE-FG03-96ER40956/A006.

\appendix
\section{Neutral meson propagator at tree level}
\label{app:GNLO}

Here we calculate the neutral meson propagator.
From ${\mcal L}_{\rm LO}$ we find
\beq
G_N^{-1} &=&G_0^{-1}+V\\
\(G_0^{-1}\)_{ab}&=&(p^2+\chi_{a})\delta_{ab}\epsilon_a\\
V_{ab}&=&\frac{m_0^2+\alpha p^2}{N}\epsilon_a\epsilon_b \,.
\eeq
The full propagator is thus
\beq
G_N= \(G_0^{-1}+V\)^{-1} = (1+G_0 V)^{-1}G_0\,.
\label{eq:Gres0}
\eeq
Because $V$ is an outer product,
the combination $G_0 V$ is proportional to a projection operator
\beq
A &\equiv& {G_0 V \over \tr(G_0 V)} \,,\quad A^2 = A \,.
\eeq
Thus for any function $f$,
\beq
f(A) - f(0) = A [f(1) - f(0)] \,,
\eeq
and so
\beq
\left(1 + G_0 V\right)^{-1}  &=& 
\left(1 + \tr(G_0 V) A \right)^{-1} \nonumber \\
&=& 1 + A\left[ {1 \over 1+ \tr(G_0 V)} - 1 \right] \nonumber \\
&=& 1 - {G_0 V \over 1 + \tr(G_0 V)} \,.
\eeq
Inserting this in Eq.~(\ref{eq:Gres0}), we find
\beq
G_N = G_0 - {G_0 V G_0 \over 1 + \tr(G_0 V)} \,.
\label{eq:Gres1}
\eeq
which reproduces the result of Ref.~\cite{BGPQ}.

The analytic structure of the propagator is not clear from this
result. In particular, its diagonal elements $[G_N]_{aa}$
appear to contain double poles (from the two factors of $G_0$ in
the second term). We know, however, that if we restrict 
ourselves to the physical sea-quark sector then
$G_N$ cannot contain double poles.
Thus there are cancelations hidden in Eq.~(\ref{eq:Gres1})
which we want to make explicit.

To do so we need to introduce the restriction
of the various matrices to the sea-sea sector.
We denote these restrictions by overbars.
We first observe that the previous steps go through identically
for the restricted matrices
\beq
\(\overline{ G_N^{-1}}\)^{-1} = \overline G_0 - 
{\overline G_0 \;\overline V \;\overline G_0 \over 
1 + \tr(\overline G_0 \overline V)} \,;
\label{eq:Gbarres1}
\eeq
Comparing this with the restriction of Eq.~(\ref{eq:Gres1}),
\beq
\overline G_N = \overline G_0 - 
{\overline G_0\; \overline V \;\overline G_0 \over 1 + \tr(G_0 V)} \,;
\label{eq:Gbarres2}
\eeq
and using the result
\beq
\tr(G_0 V) = \tr(\overline G_0 \overline V) \label{eq:restrict}\,,
\eeq
(which follows because the valence and ghost contributions cancel)
we find
\beq
\overline G_N = \(\overline{ G_N^{-1}}\)^{-1} \,.
\eeq
In other words, restriction to the sea-quark sector
commutes with inversion.
This result is non-trivial because $G_N^{-1}$ is not block diagonal---it
encapsulates the lack of feedback from the valence to the sea-quark sector.

Returning to the simplification of the propagator,
we note that
\beq
{\det\left[\overline{G}_N^{-1}\right]}/{\det\left[\overline{G}_0^{-1}\right]}
&=&\det\left[\overline{G}_0\; \overline{G}_N^{-1}\right]\nonumber\\
&=&\det\left[1+\tr(\overline G_0 \overline V)\overline{A}\right]\nonumber\\
&=&\exp \tr \ln\left[1+\tr(\overline G_0 \overline V)\overline{A}\right]\nonumber\\
&=&\exp \ln\left[1+\tr(\overline G_0 \overline V)\right]\nonumber\\
&=&1+\tr(\overline G_0 \overline V)\,.\label{eq:detrat0}
\eeq
Thus the factor multiplying the double pole in Eq.~(\ref{eq:Gres1})
can be rewritten as
\beq
\left[1+\tr(G_0 V)\right]^{-1} = 
{\det\left[\overline{G}_0^{-1}\right]}/
{\det\left[\overline{G}_N^{-1}\right]} \,.
\label{eq:detrat}
\eeq
The determinant in the numerator is simple
\beq
\det\left[\overline{G}_0^{-1}\right] = 
(p^2 + \chi_1)^{N_1} (p^2 + \chi_2)^{N_2} (p^2 + \chi_3)^{N_3} \,.
\label{eq:freedet}
\eeq
Our final task is to evaluate the determinant in the denominator.

To do this, we note that $\overline{G}_N^{-1}=\overline{G_N^{-1}}$ 
is block diagonal.
The exact $SU(N_1)\times SU(N_2) \times SU(N_3)$ flavor symmetry
implies that there are, for each sea-quark type $i$, 
$N_i-1$ flavor non-singlet neutral pions which
are eigenvectors of $\overline G_N^{-1}$.
Since $\overline V$ projects onto flavor singlet states,
the corresponding eigenvalues are those of $\overline G_0^{-1}$,
namely $(p^2 + \chi_i)$.
The non-trivial part of $\overline G_N^{-1}$, to which $V$ does contribute,
is thus a $3\times 3$ block. In QCD, with $N_i=1$, this is the entire
matrix, and describes the $\pi_0$, $\eta$ and $\eta'$. For convenience,
we use these names for general $N_i$ as well.
We denote the restriction of matrices to this subspace with double bars.
A straightforward exercise shows that
\beq
\overline{\overline{G_N^{-1}}} &=& S R^T D R S \,,\\
S &=& \mbox{diag}(1,1,\sqrt{1+\alpha}) \,,\\
D &=& \mbox{diag}(p^2 + \chi_\pi, p^2 + \chi_{\eta}, p^2 + \chi_{\eta'})
\,,
\eeq
with $R$ a rotation matrix.
$S$ rescales the singlet field so that it has a canonical kinetic term.
The meson mass-squareds are given, up to corrections of
size $\chi^2/m_0^2$, by
\beq
\chi_\pi+\chi_\eta &=& \chi_1+\chi_2+\chi_3-\bar\chi  \label{eq:chipi1}\\
\chi_\pi\chi_\eta &=& \chi_1\chi_2\chi_3 \overline{\chi^{-1}}
\label{eq:chipi2}\\
\chi_{\eta'} &=& (m_0^2 + \bar \chi )/(1+\alpha) 
\label{eq:chipi3}
\eeq
where 
\beq
\bar\chi= {1\over N} \sum_{i=1,3} N_i\chi_i \,,\qquad
\overline{\chi^{-1}}= {1\over N} \sum_{i=1,3} N_i\chi_i^{-1} \,,
\eeq
are averages over the sea sector.

Using these results we find that
\beq
\det\left[\overline{G}_N^{-1}\right] = 
(p^2\! +\! \chi_1)^{N_1-1} (p^2\!+\!\chi_2)^{N_2-1} (p^2\!+\!\chi_3)^{N_3-1}
(p^2\!+\!\chi_\pi) (p^2\!+\!\chi_{\eta}) (p^2\!+\!\chi_{\eta'}) (1\!+\!\alpha)\,.
\eeq
Inserting this and Eq.~(\ref{eq:freedet}) into 
Eqs.~(\ref{eq:detrat}) and (\ref{eq:Gres1}) gives the
result quoted in the main text, Eq.~(\ref{eq:GNtree}).
It is straightforward to generalize this result to an arbitrary
number of different sea-quark masses.

\section{One-loop calculation of $\mcal{D}$}
\label{app:DNLO}

To extract $\mcal{D}$ using Eqs.~(\ref{eq:DPQ3}) and ~(\ref{eq:DPQ1})
we need the NLO results for the charged and neutral propagators.
We do the calculation only for $N$ degenerate sea quarks, which
we denote using the label $S$ rather than $1,2,3$.

The NLO calculation of the charged propagator
was described in Sec.~\ref{sec:res},
and the expression for $M_{AA}^2$ can be obtained using Eq.~(\ref{eq:MABNLO}).
Here we also need the wavefunction renormalization factor, for which we find
\beq
{\cal Z}_A = \left[1 - {8\over f^2}(L_4 N \chi_S + L_5 \chi_A)
      - {N\over 3} {1 \over 16 \pi^2 f^2} {\chi_S+\chi_A \over 2}
\log\left({\chi_S+\chi_A \over 2} \right) \right]
\,. \label{eq:ZAres}
\eeq

For the neutral propagator we need to generalize the calculation
of app.~\ref{app:GNLO} to NLO.
The inverse propagator becomes
\beq
G_N^{-1} = G_0^{-1} + V + \Sigma \,,
\eeq
where $\Sigma$ contains the NLO tree and one-loop contributions.
The structure of $\Sigma$ is such that the method
used to calculate $G^{-1}$ in Sec.~\ref{sec:theory} 
does not apply, and we simply invert the matrix by brute force. 

For our purposes, it is sufficient to consider the restriction
of $G_N$ to the three-dimensional basis
\beq
(\pi_{AA}, \eta', \pi_{\tilde A\tilde A}) \,,\qquad
\eta' = {1 \over \sqrt{N}} \sum_{i=1}^N \pi_{ii} \,.
\eeq
This is because, first, we want only the $\pi_{AA}$ propagator
and so do not need to introduce an additional valence quark;
and, second, because we use degenerate sea quarks so that there
is no mixing of $\pi_{AA}$ with flavor non-singlet neutral sea-quark mesons.
The generalization to non-degenerate
sea quarks, which involves a larger basis of neutral states, 
is straightforward in principle, but tedious in practice,
and we have not carried it out.

The contributions to $\Sigma$ fall into two classes:
those that are common to the charged mesons $\Sigma_{C}$, 
and those that are special to the neutral mesons $\Sigma_{N}$.
The former can be obtained from the results of Sec.~\ref{sec:res},
and we find
\beq
G_0^{-1}+\Sigma_C &=& \mbox{diag}(v, w, -v) \,,\\
v &=& (p^2 + M_{AA}^2)/{\cal Z}_{A} \,,\label{eq:vdef}\\ 
w &=& (p^2 + M_{SS}^2)/{\cal Z}_{S} \,.
\eeq
Here $M_{SS}^2$ is the squared mass of flavor non-singlet mesons
composed of sea quarks evaluated at NLO, and ${\cal Z}_S$ the
corresponding wavefunction renormalization.
Their expressions can be obtained from those for $M_{AA}^2$ and
${\cal Z}_A$ by the substitution $\chi_A \to \chi_S$.

We now turn to $\Sigma_N$.
Because of the graded symmetry,
it shares a particular matrix structure with $V$, 
and it is  convenient to make this explicit
by the following definitions:
\beq
V + \Sigma_N = \left(\begin{array}{ccc}
                        x &\sqrt{N} y & -x\\
                        \sqrt{N}y & Nz & -\sqrt{N} y \\
                        -x&-\sqrt{N}y&  x \end{array} \right) \,.
\label{eq:sigmaNUform}
\eeq
The contributions to this matrix from $V$ are
\beq
x_V=y_V=z_V=x_0 = (m_0^2+\alpha p^2)/N \,.
\eeq
Thus we write $x=x_0 +\delta x$, $y= x_0 +\delta y$ and
$z= x_0+\delta z$.

\makefig{fig:DNLO}{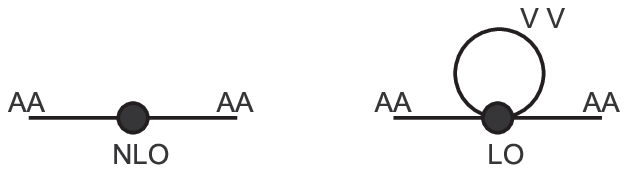}{1}{Diagrams contributing to $\Sigma_N$. 
The letters next to the lines denote the flavor indices of 
the propagating mesons. ``LO'' and ``NLO'' describe the order 
of the vertex that makes the diagram contribute at NLO.}

The diagrams contributing to $\Sigma_N$ 
are of the general form shown in Fig.~\ref{fig:DNLO}.
The tree level contributions come from the two-meson vertices 
included in $\mcal{L}_{NLO,2}$ [Eq.~(\ref{eq:chL3})]. 
The $v_2$ term acts as a subleading correction to $m_0^2$.
Since at LO [Eq.~(\ref{eq:DLO})], 
$\mcal{D}$ is independent of $m_0^2$,
it follows that at NLO $\mcal{D}$ can at most depend on its leading value.
Thus $v_2$ contributes only at NNLO.
The other tree level diagram comes 
from the $L_7$ term which gives the following contributions:
\beq
\delta^{(1)} x = {8 \over f^2} 2 L_7  \chi_A^2 \,,\quad
\delta^{(1)} y = {8 \over f^2} 2 L_7  \chi_A \chi_S \,,\quad
\delta^{(1)} z = {8 \over f^2} 2 L_7  \chi_S^2 \,.
\eeq

\makefig{fig:Qlines}{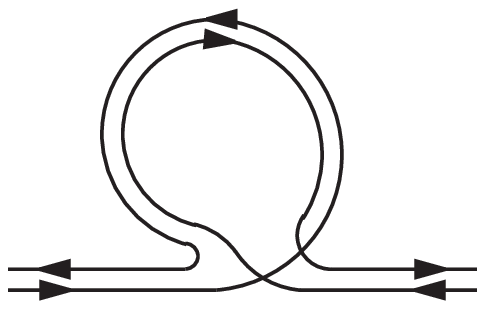}{2}{Each line in the diagram 
represents a single quark field. This is the only loop diagram 
of the type that appears in Fig.~\ref{fig:DNLO} in which the 
quark lines are disconnected.}

Figure \ref{fig:Qlines} shows the quark line 
structure of the only one-loop graph involving the lowest order vertices
which corresponds to disconnected quark lines, 
and so does not contribute to $\Sigma_C$.
The contribution to $\Sigma_N$ is
\beq
\delta^{(2)} x &=& 
{1\over 16 \pi^2 f^2}{1\over 3}(p^2-2\chi_A)\chi_A\log\chi_A \,,\\
\delta^{(2)} y &=& 
{1\over 16 \pi^2 f^2}{1\over 3}(p^2-\chi_A-\chi_S){\chi_A+\chi_S\over2}
                                \log\left({\chi_A+\chi_S\over2}\right) \,,\\
\delta^{(2)} z &=& 
{1\over 16 \pi^2 f^2}{1\over 3}(p^2-2\chi_S)\chi_S\log\chi_S \,.
\eeq

Collecting these contributions we end up with
\beq
G^{-1} = \left(\begin{array}{ccc}
                        v+x &\sqrt{N} y & -x\\
                        \sqrt{N}y & w+Nz & -\sqrt{N} y \\
                        -x&-\sqrt{N}y&  -v+x \end{array} \right) \,.
\eeq
The relevant part of the inverse is
\beq
G_{AA} = {1\over v} - {1\over v^2} {xw + N(xz-y^2) \over w+Nz} \,.
\eeq
The first term is the one-loop corrected single pole,
while the second contains the expected double pole.
Inserting this result
into the definitions Eqs.~(\ref{eq:DPQ1}) and (\ref{eq:DPQ3}),
and using Eqs.~(\ref{eq:ZAres}) and (\ref{eq:vdef}),
we can read off the required double-pole coefficient
\beq
{\cal D} &=& -{\cal Z_A} 
\left({xw + N(xz-y^2) \over w+Nz}\right)\bigg|_{p^2=-M_{AA}^2} \,.
\label{eq:Dres}
\eeq
Expanding in powers of $\chi$ we find
\beq
{\cal D} &\approx& -{\cal Z_A} 
\left\{
{w\over N}\left(1-{w\over Nx_0}\right)
+ (\delta x + \delta z - 2 \delta y)
\right\}\bigg|_{p^2=-M_{AA}^2} \,.
\eeq

Note that $(x_V z_V - y_V^2)= 0$, so that a possible 
contribution proportional to $m_0^2$ cancels. 
Substituting and rearranging, we find the answer Eq.~(\ref{eq:DNLO}).



\end{document}